\documentclass[final]{siamltex}
\usepackage{latexsym}
\newtheorem{fact} [theorem]{Fact}
\newcommand{\Com}{{\mathchoice {\setbox0=\hbox{$\displaystyle\rm
        C$}\hbox{\hbox to0pt{\kern0.4\wd0\vrule height0.9\ht0\hss}\box0}}
{\setbox0=\hbox{$\textstyle\rm C$}\hbox{\hbox
to0pt{\kern0.4\wd0\vrule height0.9\ht0\hss}\box0}}
{\setbox0=\hbox{$\scriptstyle\rm C$}\hbox{\hbox
to0pt{\kern0.4\wd0\vrule height0.9\ht0\hss}\box0}}
{\setbox0=\hbox{$\scriptscriptstyle\rm C$}\hbox{\hbox
to0pt{\kern0.4\wd0\vrule height0.9\ht0\hss}\box0}}}}

\newcommand{\N}{{\rm I\!N}}
\newcommand{\Q}{{\mathchoice {\setbox0=\hbox{$\displaystyle\rm
Q$}\hbox{\raise 0.15\ht0\hbox to0pt{\kern0.4\wd0\vrule
height0.8\ht0\hss}\box0}} {\setbox0=\hbox{$\textstyle\rm
Q$}\hbox{\raise 0.15\ht0\hbox to0pt{\kern0.4\wd0\vrule
height0.8\ht0\hss}\box0}} {\setbox0=\hbox{$\scriptstyle\rm
Q$}\hbox{\raise 0.15\ht0\hbox to0pt{\kern0.4\wd0\vrule
height0.7\ht0\hss}\box0}} {\setbox0=\hbox{$\scriptscriptstyle\rm
Q$}\hbox{\raise 0.15\ht0\hbox to0pt{\kern0.4\wd0\vrule
height0.7\ht0\hss}\box0}}}}
\newcommand{\Z}{{\mathchoice
{\hbox{$\sf\textstyle Z\kern-0.4em Z$}} {\hbox{$\sf\textstyle
Z\kern-0.4em Z$}} {\hbox{$\sf\scriptstyle Z\kern-0.3em Z$}}
{\hbox{$\sf\scriptscriptstyle Z\kern-0.2em Z$}}}}

\begin{document}

\title{One-way communication complexity and the Ne\v ciporuk lower
  bound on formula size\thanks{The results  in this paper
  have previously appeared in three conference papers
  \cite{Kl97,Kl98,Kl00} at ISAAC'97, Complexity'98, and STOC'00.}}
\author{Hartmut Klauck\thanks{Address: Department of Computer Science,
  University of Calgary, Calgary Alberta T2N 1N4, Canada. Email: klauckh@cpsc.ucalgary.ca.
  Supported by Canada's NSERC  and
  MITACS and by DFG Project KL 1470/1. Most of this work was done at Johann Wolfgang Goethe-Universit\"at Frankfurt.}
}

\maketitle

\begin{abstract} In this paper the Ne\v ciporuk method for proving lower bounds on the
  size of Boolean formulae is reformulated in terms of one-way
  communication complexity. We investigate the scenarios of
  probabilistic formulae, nondeterministic formulae, and quantum
  formulae. In all cases we can use results about one-way communication
  complexity to prove lower bounds on formula size. In the latter two
  cases we newly develop the employed communication complexity bounds.
  The main results regarding formula size are as follows: A polynomial size gap between
  probabilistic/quantum and deterministic formulae. A near-quadratic
  size gap for nondeterministic formulae with limited access to
  nondeterministic bits. A near quadratic lower bound on quantum formula size, as
  well as a polynomial separation between the sizes of quantum formulae
  with and without multiple read random inputs.
  The methods for quantum and probabilistic formulae employ
  a variant of the Ne\v ciporuk bound in terms
  of the VC-dimension. Regarding communication complexity we
  give optimal separations between one-way and two-way protocols in the
  cases of limited nondeterministic and quantum communication, and we
  show that zero-error quantum one-way communication complexity asymptotically
  equals deterministic one-way communication complexity for total
  functions.
\end{abstract}



\pagestyle{myheadings}
\thispagestyle{plain}
\markboth{H. Klauck}{Communication Complexity and the Ne\v ciporuk Method}

\section{Introduction}

One of the most important goals of complexity theory is to prove lower
bounds on the size of Boolean circuits computing some explicit
functions. Currently only linear lower bounds for this complexity
measure are known. It is well known that superlinear
lower bounds are provable, however, if we restrict the
circuits to fan-out one, i.e., if we consider Boolean formulae.
The best known technique for providing these is due to
Ne\v ciporuk \cite{N66}, see also \cite{BS90}. It applies
to Boolean formulae with arbitrary gates of fan-in two. For other
methods applying to circuits over a less general basis of gates see
e.g.~\cite{BS90}. The largest lower bounds provable with Ne\v ciporuk's
method are of the order $\Theta(n^2/\log n)$.

The complexity measure of formula size is not only interesting because
formulae are restricted circuits which are easier to handle in lower bounds, but
also because the logarithm of the formula size is asymptotically
equivalent to the circuit depth. Thus increasing the range of
lower bounds for formula size is interesting.

It has become customary to consider randomized algorithms as a
standard model of computation. While randomization can be
eliminated quite efficiently using the nonuniformity of circuits,
randomized circuits are sometimes simpler to describe and more
concise than deterministic circuits. It is natural to ask whether
we can prove lower bounds for the size of randomized formulae.

More generally, we like to consider different modes of computation
other than randomization. First we are interested in
nondeterministic formulae. It turns out that general
nondeterministic formulae are as powerful as nondeterministic
circuits, and thus intractable for lower bounds with current
techniques. But this construction relies heavily on a large
consumption of nondeterministic bits guessed by the simulating
formula, in other words such a simulation drastically increases
the length of {\it proofs} involved. So we can ask whether the
size of formulae with a limited number of nondeterministic guesses
can be lower bounded, in the spirit of research on limited
nondeterminism \cite{GLM96}.

Finally, we are interested in quantum computing. The model of quantum
formulae has been introduced by Yao in \cite{Y93}. He gives a
superlinear lower bound for quantum formulae computing the MAJORITY
function. Later Roychowdhury and Vatan \cite{RV01} proved that a
somewhat weaker form of the
classical Ne\v ciporuk method can be applied to give
lower bounds for quantum formulae of the order $\Omega(n^2/\log^2 n)$,
and that quantum formulae can actually be simulated quite efficiently
by classical Boolean circuits.

The outline of this paper is the following. First we observe that
the Ne\v ciporuk method can be defined in terms of one-way
communication complexity. While this observation is not relevant
for deterministic computations, its power becomes useful if we
consider other modes of computation. First we consider
probabilistic formulae. We derive a variation of the Ne\v ciporuk
bound in terms of randomized communication complexity and, using
results from that area, a combinatorial variant involving the
VC-dimension. Applying this lower bound we show a near-quadratic
lower bound for probabilistic formula size (corollary 3.7). We
also show that there is a function, for which probabilistic
formulae are smaller by a factor of $\sqrt{n}$ than deterministic
formulae and even Las Vegas (zero error) formulae (corollary
3.13). This is shown to be the maximal such gap provable if the
lower bound for deterministic formulae is given by the Ne\v
ciporuk method. Furthermore we observe that the standard Ne\v
ciporuk bound asymptotically also works for Las Vegas formulae.

We then introduce Ne\v ciporuk methods for nondeterministic
formulae and for quantum formulae. To apply these generalizations
we have to provide lower bounds for one-way communication
complexity with limited nondeterminism, and for quantum one-way
communication complexity. For both measures lower bounds
explicitly depending on the one-way restriction  were unknown
prior to this work. Since the communication problems we
investigate are asymmetric (i.e., Bob receives much fewer inputs
than Alice) our results show optimal separations between one- and
two- round communication complexity for limited nondeterministic
and for quantum communication complexity. Such separations have
been known previously only for deterministic and probabilistic
protocols, see \cite{KNR99,PS84}.

In the nondeterministic case we give a specific combinatorial
argument for the communication lower bound (Theorem 5.5). In the
quantum case we give a general lower bound method based on the
VC-dimension (Theorem 5.9), that can also be extended to the case
where the players share prior entanglement. Furthermore we show
that exact and Las Vegas quantum one-way communication complexity
are never much smaller than deterministic one-way communication
complexity for total functions (theorems 5.11/5.12).

Then we are ready to give Ne\v ciporuk style lower bound methods
for nondeterministic formulae and quantum formulae. In the
nondeterministic case we show that for an explicit function there
is a threshold on the amount of nondeterminism needed for
efficient formulae, i.e., a near-quadratic size gap occurs between
formulae allowed to make a certain amount of nondeterministic
guesses, and formulae allowed a logarithmic factor more. The
threshold is polynomial in the input length (Theorem 6.4).

For quantum formulae we show a lower bound of $\Omega(n^2/\log
n)$, improving on the best previously known bound given in
\cite{RV01} (Theorem 6.11). More importantly, our bound also
applies to a more general model of quantum formulae, which are
e.g.~allowed to access multiple read random variables. This
feature makes these generalized quantum formulae a proper
generalization of both quantum formulae and probabilistic
formulae. It turns out that we can give a $\Omega(\sqrt{n}/\log
n)$ separation between formulae with multiple read random
variables and without this option, even if the former are
classical and the latter are quantum (corollary 6.6). Thus quantum
formulae as defined by Yao are not capable of efficiently
simulating classical probabilistic formulae. We show that the
VC-dimension variant of the Ne\v ciporuk bound holds for
generalized quantum formulae and the standard Ne\v ciporuk bound
holds for generalized quantum Las Vegas formulae (Theorem 6.10).

The organization of the paper is as follows: in \S 2 we describe
some preliminaries regarding the VC-dimension, classical
communication complexity, and Boolean circuits. In \S 3 we expose
the basic lower bound approach and apply the idea to probabilistic
formulae. In \S 4 we give more background on quantum computing and
information theory. In \S 5 we give the lower bounds for
nondeterministic and quantum one-way communication complexity. In
\S 6 we derive our results for nondeterministic and quantum
formulae and apply those bounds. In \S 7 we give some conclusions.

\section{Preliminaries}

\subsection{The VC-dimension}

We start with a useful combinatorial concept \cite{VC71}, the
Vapnik-Chervonenkis dimension. This will be employed to derive lower
bounds for one-way communication complexity and then to give
generalizations of the Ne\v ciporuk lower bound on formula size.

\begin{definition}
A set $S$ is shattered by a set of Boolean functions $\cal F$,
if for all $R\subseteq S$ there is a function $f\in \cal F$, so that
for all $x\in S$: $f(x)=1 \iff x\in R$.

The size of a largest set shattered by $\cal F$ is called the
 VC-Dimension $VC({\cal F})$ of
$\cal F$.
\end{definition}

The following fact \cite{VC71} will be useful.

\begin{fact}
Let $\cal F$ be a set of Boolean functions $f:X\to\{0,1\}$. Then
\[2^{VC({\cal F})}\le |{\cal F}|\le (|X|+1)^{VC({\cal F})}.\]
\end{fact}

\subsection{One-way communication complexity}

We now define the model of one-way communication complexity, first
described by Yao \cite{Y79}. See \cite{KN97} for more details on
communication complexity.

\begin{definition}
Let $f: X\times Y\to \{0,1\}$ be a function. Two
players Alice and Bob with unrestricted computational power
receive inputs $x\in X, y\in Y$ to the function.

Alice sends a binary encoded message to Bob, who then computes the
function value. The complexity of a protocol is the worst case length
of the message sent (over all inputs).

The deterministic one-way communication complexity of $f$, denoted
$D(f)$, is the complexity of an optimal deterministic protocol computing $f$.

In the case Bob sends one message and Alice announces the result we
use the notation $D^B(f)$.

The communication matrix of a function $f$ is the matrix $M$ with
$M(x,y)=f(x,y)$ for all inputs $x,y$.
\end{definition}

We will consider different modes of acceptance for communication
protocols. Let us begin with nondeterminism.

\begin{definition}
In a nondeterministic one-way protocol for a Boolean function
$f:X\times Y\to\{0,1\}$ Alice first guesses nondeterministically a
sequence of $s$ bits. Then she sends a message to Bob, depending on the
sequence and her own input. Bob computes the function
value. Note that the guessed sequence is only known to Alice. An input
is accepted, if there is a guess, so that Bob accepts given the
message and his input. All
other inputs are rejected.

The complexity of a nondeterministic one-way protocol with $s$
nondeterministic bits is the length of the longest message used.

The nondeterministic communication complexity $N(f)$ is the complexity
of an optimal one-way protocol for $f$ using arbitrarily many
nondeterministic bits.

$N_s(f)$ denotes the complexity of an optimal nondeterministic
protocol for $f$, which uses at most $s$ private nondeterministic bits for every input.
\end{definition}

Note that if we do not restrict the number of nondeterministic bits,
then nondeterministic protocols with more than one round of
communication can be simulated: Alice guesses a dialogue, sends it if
it is consistent with her input, Bob checks the same with his input and accepts if
acceptance is implied by the dialogue.

While nondeterministic communication is a theoretically motivated
model, probabilistic communication is the most powerful realistic
model of communication besides quantum mechanical models.

\begin{definition}
In a probabilistic protocol with private random coins Alice and Bob
each possess a source of independent random bits with uniform
distribution. The players are allowed to access that source and
communicate depending on their inputs and the random bits they
read. We distinguish the following modes of acceptance:

\begin{remunerate}
\item In a Las Vegas protocol the players are not allowed to err. They
  may, however, give up without an output with some probability
  $\epsilon$. The complexity of a one-way protocol is the worst case length of
  a message used by the protocol, the Las Vegas complexity of a
  function $f$ is the complexity of an optimal Las Vegas protocol
  computing $f$, and is denoted $R_{0,\epsilon}(f)$.
\item In a probabilistic protocol with bounded error $\epsilon$
the output has to be correct with probability at least
$1-\epsilon$. The complexity of a protocol is the worst case length of
the message sent (over all inputs and the random guesses), the
complexity of a function is the complexity of an optimal protocol
computing that function and is denoted $R_\epsilon(f)$. For
$\epsilon=1/3$ the notation is abbreviated to $R(f)$.
\item A bounded error protocol is a Monte Carlo protocol, if inputs
  with $f(x_A,x_B)=0$ are rejected with certainty.
\end{remunerate}

We also consider probabilistic communication with public
randomness. Here the players have access to a shared source of random
bits without communicating. Complexity in this model is denoted
$R^{pub}$, with acceptance defined as above.
\end{definition}

The difference between probabilistic communication complexity with public and with private
random bits is actually only an additive $O(\log n)$ as shown in
\cite{Ne91} by an argument based on the nonuniformity of the model.

The following communication problems are frequently considered in the
literature about communication complexity.

\begin{definition} Disjointness problem\\
$DISJ_n(x_1\ldots x_n,y_1\ldots y_n)=1$ $\iff \forall i:\neg
x_i\vee\neg y_i$. The function accepts, if the two sets described by
the inputs are disjoint.

Index function\\
$IX_{2^n}(x_1\ldots x_{2^n},y_1\ldots y_n)=1\iff x_y=1$.
\end{definition}

The deterministic one-way communication complexity of a function
can be characterized as follows. Let $row(f)$ be the
number of different rows in the communication matrix of $f$.

\begin{fact}
$D(f)=\lceil \log row(f)\rceil$.
\end{fact}

It is relatively easy to estimate the deterministic
one-way communication complexity using this fact. As an example
consider the index function, note that obviously $D^{B}(IX_n)=\log n$. It is easy to see
with Fact 2.7 that $D(IX_n)=n$, since there are
$2^n$ different rows in the communication matrix of $IX_n$.
In \cite{KNR99} it is shown that also $R^{pub}(IX_n)=\Omega(n)$.

A general lower bound method for probabilistic one-way communication
complexity is shown in \cite{KNR99}.

We consider the VC-dimension for functions as follows.

\begin{definition} For a function $f:X\times Y\to\{0,1\}$ let
${\cal F}=\{g|\exists x\in X:\forall y\in Y:g(y)=f(x,y)\}$. Then
 define $VC(f)=VC({\cal F})$.\end{definition}

\begin{fact}
$R^{pub}(f)=\Omega(VC(f))$
\end{fact}

In \S 5.2 we will generalize this result to quantum one-way protocols.

With the above definition $\lceil\log|{\cal F}|\rceil=D(f)$.
Then $VC(f)\le D(f)\le\lceil\log (|Y|+1)\cdot VC(f)\rceil$ due
to Fact 2.2.

Las Vegas communication can be quadratically more efficient than
deterministic communication in many-round protocols for total
functions \cite{KN97}. For one-way protocols the situation is different
\cite{HS01}.

\begin{fact} For all total functions $f$:\\
$R^{pub}_{0,1/2}(f)\ge D(f)/2$.
\end{fact}

We will also generalize this result to quantum communication in
\S 5.2. In our proofs for these generalizations we will employ quantum
information theoretic methods as opposed to the proofs in the
classical case, which were relying on combinatorial techniques.

\subsection{Circuits and formulae}

We now define the models of Boolean circuits and formulae. Note that
we do not consider questions  of uniformity of families of such
circuits.
For the definition of a Boolean circuit we refer to
  \cite{BS90}. We consider circuits with fan-in 2.
While it is well known that almost all  $f:\{0,1\}^n\to\{0,1\}$
need circuit size $\Theta(2^n/n)$ (see e.g.~\cite{BS90}), superlinear
lower bounds for explicit functions are only known for restricted
models of circuits.

\begin{definition}
A (deterministic) Boolean formula is a Boolean circuit with fan-in 2 and fan-out
1. The Boolean inputs may be read arbitrarily often, the gates are
arbitrary, constants 0,1 may be read.

The size (or length) of a deterministic Boolean formula is the number
of its nonconstant leaves.
\end{definition}

It is possible to show that for Boolean functions the logarithm of the
formula size is linearly related to the optimal circuit depth  (see \cite{BS90}).

Probabilistic formulae have been considered in
\cite{V84,B85,DZ97} with the purpose of constructing efficient (deterministic)
monotone formulae for the majority function in a probabilistic manner.

The ordinary model of a probabilistic formula is a probability
distribution on deterministic formulae. Since formulae are also an
interesting datastructure we are interested in a more compact
model. ``Fair'' probabilistic formulae are formulae that read input
variables plus additional random variables. The other model will be
called ``strong'' probabilistic formulae.

\begin{definition}
A fair probabilistic formula is a Boolean formula, which works on
input variables and additional random variables  $r_1,\ldots,r_m$, a
strong probabilistic formula is a probability distribution $F$ on
deterministic Boolean formulae. Fair resp.~strong probabilistic formulae
$F$ compute a Boolean function $f$ with bounded error, if

\[\Pr[F(x)\neq f(x)]\le 1/3.\]

Fair resp.~strong probabilistic formulae $F$ are Monte Carlo formulae for $f$
(i.e., have one-sided error), if

\[\Pr[F(x)=0| f(x)=1]\le 1/2 \mbox{ and } \Pr[F(x)=1| f(x)=0]=0.\]

A Las Vegas formula consists of 2 Boolean formula. One formula
computes the output, the other (verifying) formula indicates whether
the computation of the first can be trusted or not. Both work on the
same inputs. There are four different outputs, of which two are
interpreted as ``?'' (the verifying formula rejects), and the other as
0 resp.~1. A Las Vegas formula $F$ computes $f$, if the outputs 0 and
1 are always correct, and

\[\Pr[F(x)=?]\le 1/2.\]

The size of a fair probabilistic formula is the number of its
nonconstant leaves, the size of a strong probabilistic formula is the
expected size of a deterministic formula according to $F$.
\end{definition}

It is easy to see that one can decrease the error probability to
arbitrarily small constants, while increasing the size by a constant
factor, therefore we will sometimes allow different error probabilities.

A strong probabilistic formula $F$ can be transformed into a deterministic
formula. For Monte Carlo formulae this increases the size by a factor
of $O(n)$: choose $O(n)$ formulae randomly according to $F$ and
connect them by an OR gate. An application of the Chernov inequality
proves that the error probability is so small that no errors are
possible anymore. Strong formulae with bounded (two-sided) error are
derandomized by picking $O(n)$ formulae and connecting them by an
approximative majority function. That function outputs 1 on $n$
Boolean variables if at least $2n/3$ have the value 1, and outputs 0,
if at most $n/3$ variables have the value 1. An approximative
majority function can be computed by a deterministic formula of size
$O(n^2)$, see \cite{V84, B85}. Thus the size increases by a
factor of $O(n^2)$.

Let us remark that strong probabilistic formulae may have sublinear
length, this is impossible for fair probabilistic formulae depending
on all inputs. An approximative majority function may be computed by a
strong probabilistic formula through picking a random input and
outputting its value.

We will later also consider nondeterministic formulae.

\begin{definition}
A nondeterministic formula with $s$ nondeterministic bits is a formula
with additional input variables $a_1,\ldots,a_s$.
The formula accepts an input $x$, if there is a setting of the
variables $a$, so that $(a,x)$ is accepted.
\end{definition}

\section{The general lower bound method and probabilistic formulae}

There are some well known results giving lower bounds for the length
of Boolean formulae. The method of Ne\v ciporuk \cite{N66,BS90}
remains the one giving
the largest lower bounds among those methods working
for formulae in which all fan-in 2 functions are allowed as gates. For
other methods see \cite{BS90} and \cite{BNS92}; a characterization for
formula size with gates AND, OR, NOT using the communication
complexity of a certain game is also known (see
\cite{KN97}). For such formulae the largest known lower bound is a
near-cubic bound due to H\aa stad \cite{H98}.

Let us first give the standard definition of the Ne\v ciporuk bound.

Let $f$ be a function on the $n$ variables in
$X=\{x_1,\ldots,x_n\}$. For a subset
$S\subseteq X$ let a subfunction on $S$ be a function induced by $f$
by fixing the variables in $X-S$. The set of all subfunctions on $S$
is called the set of $S$-subfunctions of $f$.

\begin{fact} [Ne\v ciporuk]
Let $f$ be a Boolean function on $n$ variables. Let $S_1,\ldots, S_k$
be a partition of the variables and $s_i$ the number of
$S_i$-subfunctions on $f$. Then every deterministic Boolean formulae
for $f$ has size at least \[(1/4)\sum_{i=1}^k \log s_i.\]
\end{fact}

It is easy to see that the Ne\v ciporuk function $(1/4)\sum_{i=1}^k
\log s_i$ is never larger than $n^2/\log n$.

\begin{definition} The function "indirect storage access"\ ISA is
defined as follows: there are three blocks of inputs $U,X,Y$ with
$|U|=\log n-\log\log n$, $|X|=|Y|=n$. $U$ addresses a block of length $\log n$
in $X$, which addresses a bit in $Y$. This bit is the output, thus
$ISA(U,X,Y)=Y_{X_U}$.
\end{definition}

The following is proved e.g.~in \cite{BS90},\cite{Z95}.

\begin{fact}
Every deterministic formula for ISA has size $\Omega(n^2/\log n)$.

There is a deterministic formula for ISA with size $O(n^2/\log n)$.
\end{fact}

We are now going to generalize the Ne\v ciporuk method to
probabilistic formulae, and later to nondeterministic and quantum
formulae. We will use a simple connection to one-way communication
complexity and use the guidance obtained by this connection to give
lower bounds from lower bounds in communication complexity. In the
case of probabilistic formulae we will employ the VC-dimension to give
lower bounds. Informally speaking we will replace the log of the size
of the set of subfunctions by the VC-dimension of that set and get a
lower bound for probabilistic formulae.

Our lower bounds are valid in the model of strong probabilistic
formulae. Corollary 3.7 shows that even strong probabilistic formulae with
two-sided error do
not help to decrease the size of formulae for $ISA$. All upper bounds will be
given for fair formulae.

We are going to show that the (standard) Ne\v ciporuk is at most
a factor of $O(\sqrt{n})$ larger than the probabilistic formula size
for total functions. Thus the maximal gap we can show using the
currently best general lower bound method is limited.

On the other hand we describe a Boolean function, for which fair
probabilistic formulae with one-sided error are a factor $\Theta(\sqrt
n)$ smaller than Las Vegas formulae, as well as a similar gap between
one-sided error formulae and two-sided error formulae.
The lower bound on Las Vegas formulae uses the new observation that
the standard Ne\v ciporuk bound asymptotically also works for Las Vegas formulae.

\subsection{Lower bounds for probabilistic formulae}

We now derive a Ne\v ciporuk type bound with one-way communication.

\begin{definition}
Let $f$ be a Boolean function on $n$ inputs and let
$y_1\ldots y_k$ be a partition of the input variables.

We consider $k$ communication problems for $i=1,\ldots,k$. Player Bob
receives all inputs in $y_i$, player Alice receives all other inputs.
The deterministic one-way communication complexity of $f$ under this partition
of inputs is called $D(f_i)$. The public coin bounded error
one-way communication complexity of $f$ under this partition
of inputs is called $R^{pub}(f_i)$.

The probabilistic Ne\v ciporuk function is $(1/4)\sum_i R^{pub}(f_i)$.
\end{definition}

It is easy to see that $(1/4)\sum_i D(f_i)$ coincides with the
standard Ne\v ciporuk function and is therefore a lower bound for
deterministic formula size due to Fact 3.1.

\begin{theorem} The probabilistic Ne\v ciporuk function is a lower
  bound for the size of strong probabilistic formulae with bounded error.
\end{theorem}

\begin{proof}
We will show for every partition $y_1,\ldots,y_k$ of the inputs, how a
strong probabilistic formula $F$ can be simulated in the $k$
communication games. Let $F_i$ be the
distribution over deterministic formulae on variables in $y_i$ induced
by picking a deterministic formula as in $F$ and restricting to the
subformula with all leaves labeled by variables in $y_i$ and
containing all paths from these to the root. We
want to simulate the formula in game $i$ so that the probabilistic
one-way communication is bounded by the expected number of leaves in $F_i$.

We are given a probabilistic formula $F$. The players now pick a
deterministic formula $F'$ induced by $F$ with their public random
bits, Player Alice knows all the inputs except those in $y_i$. This
also fixes a subformula $F'_i$ drawn from $F_i$. Actually the players
have only access to an arbitrarily large public random string, so the
distributions $F_i$ may only be approximated within arbitrary
precision. This alters success probabilities by arbitrary small
values. We disregard these marginal probability changes.

Let $V_i$ contain the vertices in $F'_i$, which have 2 predecessors in
$F'_i$, and let $P_i$ contain all paths, which start in $V_i$ or at a
leaf, and which end in $V_i$ or at the root, but contain no further
vertices from $V_i$. It suffices, if Alice sends 2 bits for each such
path, which shows, whether the last gate of the path computes $0,1,g$,
or $\neg g$, for the function $g$ computed by the first gate of the
path. Then Bob can evaluate the formula alone.

There are at most $2|V_i|+1$ paths as described, since the fan-in of
the formula is 2. Thus the overall communication is $4|V_i|+2$. The
set of leaves $L_i$ with variables from $y_i$ has $|V_i|+1$ elements,
and thus
\[R^{pub}(f_i)\le 4|V_i|+2<4|L_i|\] and $1/4\sum_i
R^{pub}(f_i)$ is a lower bound for the length $E[\sum_i
|L_i|]=\sum_i E[|L_i|]$ of the probabilistic formula.
\qquad\end{proof}

Let $VC(f_i)$ denote the VC-dimension of the communication problem $f_i$.
We call $\sum_i VC(f_i)$ the VC-Ne\v ciporuk function.

\begin{corollary}
The VC-Ne\v ciporuk function is an asymptotical lower bound for the
length of strong probabilistic formulae with bounded error.

The standard Ne\v ciporuk function is an asymptotical lower bound for
the length of strong Las Vegas formulae for total functions.
\end{corollary}

\begin{proof} Using Fact 2.9 the VC-dimension is an asymptotical lower bound
for the probabilistic public coin bounded error
one-way communication complexity.

As in the proof of Theorem 3.5 we may simulate a Las Vegas formula by
Las Vegas public coin one-way protocols.
Using Fact 2.10 public coin Las Vegas one-way protocols for total functions can only
be a constant factor more efficient than optimal deterministic one-way
protocols.\qquad\end{proof}

According to Fact 3.3 the deterministic formula length of the indirect
storage access function  (ISA) from definition 3.2 is $\Theta(n^2/\log n)$.
We now employ our method to show a lower bound of the same order for
strong bounded error probabilistic formulae. Thus ISA is an explicit
function for which strong probabilism does not allow to decrease
formula size significantly.

\begin{corollary}
Every strong probabilistic formula for the ISA function (with
bounded error) has length $\Omega(n^2/\log n)$.
\end{corollary}

\begin{proof} $ISA$ has inputs $Y,X,U$ and computes $Y_{X_U}$.
First we define a partition. We partition the inputs in $X$ into
$n/\log n$ blocks containing $\log n$ bits each, all other inputs
are in one additional block. In a communication game Alice receives
thus all inputs but those in one block of $X$. Let $S$ denote the set
of possible values of the variables in that block. This set is
shattered: Let $R\subseteq S$ and $R=\{r_1,\ldots,r_m\}$. Then set the
pointer $U$ to the block of inputs belonging to Bob, and set $Y_i=1\iff i\in R$.

Thus the  VC-dimension of $f_i$ is at least $|S|=n$. Since there are
$n/\log n$ communication games, the result follows.
\qquad\end{proof}

The next result would be trivial for deterministic or for fair
probabilistic formulae, but strong probabilistic formulae can compute
functions depending on all inputs in sublinear size. Consider e.g.~the
approximate majority function. This partial function can be
computed by a strong probabilistic formulae of length 1 by picking a
random input variable. For total functions on the other hand we have:

\begin{corollary}
Every strong probabilistic formula, which computes a total function
depending on $n$ variables has length $\Omega(n)$.
\end{corollary}

\begin{proof} We partition the inputs into $n$ blocks containing one
variable each. In a communication game Alice receives thus $n-1$
variables, and Bob receives 1 variable. Since the function depends on
both Alice's and Bob's inputs, the deterministic communication
complexity is at least 1. If the probabilistic one-way communication
were 0, the error would be 1/2, thus the protocol would not compute correctly.
\qquad\end{proof}

Fact 2.2 shows that for a function $f:X\times Y\to\{0,1\}$ it is true that
$D(f)\le\lceil VC(f)\cdot\log(|Y|+1)\rceil$. This leads to

\begin{theorem}
For all total functions $f:\{0,1\}^n\to\{0,1\}$ having  a strong
probabilistic formula of length $s$, and for all partitions  of the
inputs of $f$:
\[\frac{\sum D(f_i)}{s}=O(\sqrt{n}).\]
\end{theorem}

\begin{proof} Obviously $D(f_i)\le n$ for all $i$. Since a
partition of the inputs can contain at most $\sqrt{n}$ blocks with
more than $\sqrt{n}$ variables, these contribute at most $n\sqrt{n}$
to the Ne\v ciporuk function $\sum
D(f_i)$. All smaller blocks satisfy
$D(f_i)\le\lceil\sqrt{n}\cdot VC(f_i)\rceil$. Thus overall
$\sum D(f_i)\le O(\sqrt{n}(n+\sum VC(f_i)))=O(\sqrt{n}s)$, with
corollary 3.8 and Theorem 3.5.\qquad\end{proof}

If a total function has an efficient (say linear length) probabilistic
formula, then the  Ne\v ciporuk method does not give near-quadratic lower
bounds.

\subsection{A function, for which Monte Carlo probabilism helps}

We now describe a function, for which Monte Carlo probabilism
helps  as much as we can possibly show under the constraint that
the lower bound for deterministic formulae is given using the Ne\v
ciporuk method. We find such a complexity gap even between strong
Las Vegas formulae and fair Monte Carlo formulae.

\begin{definition}
The matrix product function $MP$ receives two $n\times n$-matrices
$T^{(1)},T^{(2)}$ over $\Z_2$ as input and accepts if and only if
their product is not the all zero matrix.
\end{definition}

\begin{theorem} The $MP$ function can be computed by a fair Monte
  Carlo formula of length $O(n^2)$.
\end{theorem}

\begin{proof} We use a fingerprinting technique similar to the one used
in matrix product verification \cite{MR95}, but  adapted to be computable
by a formula.
First we construct a vector as a fingerprint for each
matrix using some random input variables. Then we multiply the
fingerprints and obtain a bit. This bit is always zero, if the matrix
product is zero, otherwise it is 1 with probability 1/4. Thus we
obtain a Monte Carlo formula.

Let $r^{(1)},r^{(2)}$ be random strings of $n$ bits each. The
fingerprints are defined as
\[F^{(1)}[k]=\bigoplus_{i=1}^n r^{(1)} [i]T^{(1)}[i,k]
\mbox{ and } F^{(2)}[k]=\bigoplus_{j=1}^n T^{(2)}[k,j]r^{(2)}[j].\]
Then let \[b=\bigoplus_{k=1}^n F^{(1)}[k]\wedge F^{(2)}[k].\]
Obviously $b$ can be computed by a formula of linear length.

Assume $T^{(1)}T^{(2)}=0$. Then $b=r^{(1)}T^{(1)}T^{(2)}r^{(2)}=0$ for
all $r^{(1)}$ and $r^{(2)}$.

If on the other hand $T^{(1)}T^{(2)}\neq 0$, then $i,j$ exist such that $\bigoplus_k
T^{(1)}[i,k]T^{(2)}[k,j]=1$. Fix all random bits except $r^{(1)}[i]$
and $r^{(2)}[j]$ arbitrarily. Note that
\[b=\bigoplus_{i,j=1}^n\left(r^{(1)}[i]r^{(2)}[j]\cdot\bigoplus_{k=1}^n
  T^{(1)}[i,k]T^{(2)}[k,j]\right).\]
Regardless how the values of sums for other $i,j$ look, one of the values of
$r^{(1)}[i]$ and $r^{(2)}[j]$ yields the result $b=1$, this happens with
probability $1/4$.
\qquad\end{proof}

\begin{theorem} For the $MP$ function a lower bound of $\Omega(n^3)$
holds for the length of strong Las Vegas formulae.
\end{theorem}

\begin{proof} We use the Ne\v ciporuk method. First the partition of
the inputs has to be defined. There are $n$ blocks $b_j$ with the bits
$T^{(2)}(i,j)$ for $i=1,\ldots,n$ plus one block for the remaining inputs. Then Alice
receives all inputs except $n$ bits in column $j$ of the second matrix, i.e.,
$T^{(2)}(\cdot ,j)$,
which go to Bob. We show that $MP$ has now one-way
communication complexity $\Omega(n^2)$. The Ne\v ciporuk method then
gives us a lower bound of $\Omega(n^3)$ for the length of
deterministic and strong Las Vegas formulae. W.l.o.g.~assume Bob has the bits
$T^{(2)}(i,1)$.

We construct a set of assignments to the input variables of Alice. Let
$U$ be a subspace of $\Z_2^n$ and $T_U$ be a matrix with $T_Ux=0\iff
x\in U$. For every $U$ we choose $T_U$ as $T^{(1)}$ and $T^{(2)}(i,j)=0$
for all $i$ and for $j\ge 2$. If there are $2^{\Omega(n^2)}$ pairwise
different subspaces, then we get that many different inputs.
But these inputs correspond to different rows in the
communication matrix, since all $T^{(1)}$ have different kernels. Thus
with corollary 3.6 the Las Vegas one-way communication
is $\Omega(n^2)$.

To see that there are $2^{\Omega(n^2)}$ pairwise different subspaces
of $\Z_2^n$ we count the subspaces with dimension at most $n/2$. There
are $2^n$ vectors. There are $2^n\choose n/2$ possibilities to choose
a set of $n/2$ pairwise different vectors.
Each such set generates a subspace of dimension at most $n/2$. Each
such subspace is generated by at most $2^{n/2}\choose n/2$ sets of
$n/2$ pairwise different vectors from the subspace. Hence this number
is an upper bound on the number of times a subspace is counted and
there are at least \[\frac{{2^n\choose n/2}}{{2^{n/2}\choose n/2}}\ge
2^{\Omega(n^2)}\] pairwise different subspaces of $\Z_2^n$.
\qquad\end{proof}

\begin{corollary} There is a function, that can be computed by a fair
 Monte Carlo formula of length $O(N)$, while every strong Las Vegas
 formula needs length
 $\Omega(N^{3/2})$ for this task, i.e.,
 there is a size gap of
 $\Omega(N^{1/2})$ between Las Vegas and Monte Carlo formulae.

 There is also a size gap of
 $\Omega(N^{1/2})$ between Monte Carlo formulae and
 bounded error probabilistic formulae.
\end{corollary}

\begin{proof} The first statement is proved in the previous
theorems. For the second statement we consider the following function
with 4 matrices as input. The function is the parity of the $MP$
function on the first two matrices and the complement of $MP$ on the
other two matrices.

A fair probabilistic formula can compute the function obviously with
length $O(n^2)$ following the construction in Theorem 3.11.
Assume we have a Monte Carlo formula, then fix the
first two input matrices once in a way so that their product is the 0
matrix, and then so that their product is something else. In this way
one gets Monte Carlo formulae for both $MP$ and its complement.
Then one can use both formulae on the same input and combine their
results to get a Las Vegas formula, which leads to the desired lower
bound with Theorem 3.12.

For the construction of a Las Vegas formula let $F$ be the Monte Carlo
formula for $MP$ and $G$ be the Monte Carlo formula for $\neg
MP$. Then $F$ and $\neg G$ are formulae for $MP$, so that $F$ never
erroneously accepts and is correct with probability 1/2, and $\neg G$
never erroneously rejects and is correct with probability
1/2. Assuming the function value is 0, then $F$ rejects. With
probability 1/2 also $\neg G$ rejects, otherwise we may give
up. Assuming the function value is 1, then $\neg G$ accepts. With
probability 1/2 also $F$ accepts, otherwise we may give up. The other
way round, if both formulae accept or both reject we can safely use
this result, and this result comes up with probability 1/2, the only
other possible result is that $F$ rejects and $\neg G$ accepts, in
this case we have to give up.\qquad\end{proof}

The formula described in the proof of Theorem  3.11 has the
interesting property that each input is read exactly once, while
the random inputs are read often. $MP$ cannot be computed by a
deterministic formula reading the inputs only once, since this
contradicts the size bound of Theorem 3.12. Later we will show
that $MP$ cannot be computed substantially more efficient by a
fair probabilistic formula reading its random inputs only once
than by deterministic formulae. This follows from a lower bound
for the size of such formulae given by the Ne\v ciporuk function
divided by $\log n$ (corollary 6.7). For $MP$ read-once random
inputs are practically useless.

\section{Background on quantum computing and information}

In this section we define more technical notions and describe
results we will need. We start with information theory, then
define the model of quantum formulae and give results from quantum
information theory. We also discuss programmable quantum gates.
These results are used in the following section to give lower
bounds for one-way communication complexity. Then we proceed to
apply these to derive more formula size bounds.

\subsection{Information theory}

We now define a few notions from classical information theory, see e.g.~\cite{CT91}.

\begin{definition}
Let $X$ be a random variable with values $S=\{x_1,\ldots,x_n\}$.

The entropy of $X$ is $H(X)=-\sum_{x\in S} \Pr(X=x)\log \Pr(X=x)$.

The entropy of $X$ given an event $E$ is\\
$H(X|E)=-\sum_{x\in S} \Pr(X=x|E)\log \Pr(X=x|E)$.

The conditional entropy of $X$ given a random variable $Y$ is\\
$H(X|Y)=\sum_y\Pr(Y=y)H(X|Y=y)$, where the sum is over the values of
$Y$. Note that $H(X|Y)=H(XY)-H(Y)$.

The information between $X$ and $Y$ is $H(X:Y)=H(X)-H(X|Y)$.

The conditional information between $X$ and $Y$, given $Z$, is\\
$H(X:Y|Z)=H(XZ)+H(YZ)-H(Z)-H(XYZ)$.

For $\alpha\in [0,1]$ we define $H(\alpha)=-\alpha\log \alpha-(1-\alpha)\log(1-\alpha)$.

All of the above definitions use the convention $0\log 0=0$.
\end{definition}

The following result is a simplified version of Fano's inequality, see
\cite{CT91}.

\begin{fact}
If $X,Y$ are Boolean random variables with $\Pr(X\neq Y)\le \epsilon$,
then\\
$H(X:Y)\ge H(X)-H(\epsilon)$.
\end{fact}

\begin{proof}
Let $Z=1\iff X=Y$ and $Z=0\iff X\neq Y$. Then $H(X|Y)=H(XY)-H(Y)=H(ZY)-H(Y)\le H(Z)\le H(\epsilon)$.
\qquad\end{proof}

The next lemma is similar in the sense of a "Las Vegas variant".

\begin{lemma}
Let $X$ be a random variable with a finite range of values $S$ and let
$Y$ be a random variable with range $S\cup\{x_?\}$, so that
$\Pr(Y=x|X=x)\ge 1-\epsilon$ for all $x\in S$, $\Pr(Y=x|X\neq x)=0$
for all $x\neq x_?$ and $\Pr(Y=x_?|X=x)\le\epsilon$ for all $x\in
S$. Then
$H(X:Y)\ge (1-\epsilon)H(X)$.
\end{lemma}

{\em Proof}. $H(X:Y)=H(X)-H(X|Y)$. Let $\delta=\Pr(Y=x_?)\le\epsilon$
and $\epsilon_x=\Pr(Y=x_?|X=x)\le\epsilon$ and $p_x=\Pr(X=x)$.
\begin{eqnarray*}
H(X|Y)& &\le (1-\delta )H(X|Y\neq x_?)+\delta H(X|Y=x_?)\\
& &=\delta H(X|Y=x_?)\\
& &=-\delta \sum_x \Pr(X=x|Y=x_?)\log(\Pr(X=x|Y=x_?))\\
& &=-\delta \sum_x (\epsilon_xp_x/\delta)\log(\epsilon_xp_x/\delta)\\
& &\le-\epsilon\sum_x p_x\log p_x+\delta\sum_x(\epsilon_xp_x/\delta)\log(\delta/\epsilon_x)\\
& &\le\epsilon H(X)+\delta\log \sum_x p_x \mbox{ with Jensen's
  inequality}\\
& &\le\epsilon H(X).\qquad\endproof
\end{eqnarray*}

\subsection{Quantum computation}

We refer to \cite{NC00} for a thorough introduction into the field.
Let us briefly mention that pure quantum states are unit vectors in a
Hilbert space written $|\psi\rangle$, inner products are denoted
$\langle \psi|\phi\rangle$, and the standard norm is
$\|\,|\psi\rangle\,\|=\sqrt{\langle \psi|\psi\rangle}$.
Outer products $|\psi\rangle\langle\phi|$ are matrix valued.

In the space $\Com^4$ we will not only consider the standard basis
$\{|00\rangle, |01\rangle, |10\rangle, |11\rangle\}$, but also the
Bell basis consisting of
\[|\Phi^+\rangle=\frac{1}{\sqrt{2}}(|00\rangle+|11\rangle),\,
|\Phi^-\rangle=\frac{1}{\sqrt{2}}(|00\rangle-|11\rangle),\]

\[|\Psi^+\rangle=\frac{1}{\sqrt{2}}(|01\rangle+|10\rangle),\,
|\Psi^-\rangle=\frac{1}{\sqrt{2}}(|01\rangle-|10\rangle).\]

The dynamics of a discrete time quantum system is described by unitary
operations. A very useful operation is the Hadamard transform.

\[H_2=\frac{1}{\sqrt{2}}\left(\begin{array}{rr}
1&1\\
1&-1\end{array}\right).\]

Then $H_n=\underbrace{H_2\otimes\cdots\otimes H_2}_n$, is the $n$-wise
tensor product of $H_2$.

The XOR operation is defined by
$XOR:|x,y\rangle\to |x,x\oplus y\rangle$ on Boolean values $x,y$.

Furthermore measurements are fundamental operations. Measuring as
well as tracing out subsystems leads to probabilistic mixtures of pure states.

\begin{definition}
An ensemble of pure states is a set $\{(p_i,|\phi_i\rangle)|1\le i\le
k\}$. Here the $p_i$ are the probabilities of the pure states
$|\phi_i\rangle$. Such an ensemble is called a mixed state.

The density matrix of a pure state $|\phi\rangle$ is the matrix
$|\phi\rangle\langle\phi|$, the density matrix of a mixed state
$\{(p_i,|\phi_i\rangle)|1\le i\le k\}$ is
\[\sum_{i=1}^k p_i|\phi_i\rangle\langle\phi_i|.\]
\end{definition}

A density matrix is always Hermitian, positive semidefinite, and has
trace 1. Thus a density matrix has nonnegative eigenvalues that sum
to 1. The results of all measurements of a mixed state are determined
by the density matrix.

A pure state in a Hilbert space $H=H_A\otimes H_B$ cannot in general be expressed
as a tensor product of pure states in the subsystems.

\begin{definition}
A mixed state $\{(p_i,|\phi_i\rangle)|1\le i\le k\}$ in
a Hilbert space $H_1\otimes H_2$ is called separable, if it has the
same density matrix as a mixed state $\{(q_i,
|\psi^1_i\rangle\otimes|\psi^2_i\rangle)|i=1,\ldots,k'\}$ for pure
states $|\psi^1_i\rangle$ from $H_1$ and $|\psi^2_i\rangle$ from
$H_2$ with $\sum_iq_i=1$ and $q_i\ge0$.
Otherwise the state is called entangled.
\end{definition}

Consider e.g.~the state
$|\Phi^+\rangle=\frac{1}{\sqrt{2}}(|00\rangle+|11\rangle)$ in $\Com^2\otimes
\Com^2$. The state is entangled and is usually called an EPR-pair.
This name refers to Einstein,
Podolsky, and Rosen, who first considered such states \cite{EPR35}.

Linear transformation on density matrices are called
superoperators. Not all superoperators are physically allowed.

\begin{definition}
  A superoperator $T$ is positive, if it sends positive semidefinite
Hermitian matrices to positive semidefinite Hermitian matrices.
A superoperator is trace preserving, if it maps matrices with trace 1
to matrices with trace 1.

A superoperator $T$ is completely positive, if every
superoperator $T\otimes I_F$ is positive, where $I_F$ is the identity
superoperator on a finite dimensional extensional $F$ of the underlying
Hilbert space.

A superoperator is physically allowed, iff it is
completely positive and trace preserving.\end{definition}

The following theorem (called Kraus representation theorem)
characterizes physically allowed superoperators in terms of unitary
operation, adding qubits, and tracing out \cite{NC00}.

\begin{fact}
The following statements are equivalent:
\begin{remunerate}
\item A superoperator $T$ sending density matrices over a Hilbert
  space $H_1$ to
  density matrices over a Hilbert space $H_2$ is physically allowed.
\item There is a Hilbert space $H_3$ with $dim(H_3)\le dim(H_1)$ and a
  unitary map $U$, so that for all density  matrices $\rho$ over $H_1$:\[T\rho=
trace_{H_1\otimes H_3}[U(\rho\otimes|0_{H_3\otimes
  H_2}\rangle\langle0_{H_3\otimes H_2}|)U^\dagger ].\]
\end{remunerate}
\end{fact}

\subsection{Quantum information theory}

In this section we describe notions and results from quantum
information theory.

\begin{definition}
The von Neumann entropy of a density matrix $\rho_X$ is
$S(X)=S(\rho_X)=-trace(\rho_X\log\rho_X)$.

The conditional von Neumann entropy $S(X|Y)$ of a bipartite system
with density matrix $\rho_{XY}$ is defined as $S(XY)-S(Y)$, where the
state $\rho_Y$ of the $Y$ system is the result of a partial trace over
$X$.

The von Neumann information between two parts of a bipartite system in
a state $\rho_{XY}$ is $S(X:Y)=S(X)+S(Y)-S(XY)$ ($\rho_X$ and $\rho_Y$
are the results of partial traces).

The conditional von Neumann information of a system in state
$\rho_{XYZ}$ is $S(X:Y|Z)=S(XZ)+S(YZ)-S(Z)-S(XYZ)$.

Let ${\cal E}=\{(p_i,\rho_i)|i=1,\ldots,k\}$ be an ensemble of density
matrices. The Holevo information of the ensemble is $\chi({\cal
  E})=S(\sum_ip_i\rho_i)-\sum_ip_iS(\rho_i)$.
\end{definition}

The von Neumann entropy of a density matrix depends on the
eigenvalues only, so it is invariant under unitary transformations. If the
underlying Hilbert space has dimension $d$, then the von Neumann
entropy of a density matrix is bounded by $\log d$. A fundamental
result is the so-called Holevo bound \cite{H73}, which states an upper
bound on the amount of classical information in a quantum state.

\begin{fact}
Let $X$ be a classical random variable  with $\Pr(X=x)=p_x$. Assume
for each $x$ a quantum state with density matrix $\rho_x$ is prepared,
i.e., there is an ensemble ${\cal
  E}=\{(p_x,\rho_x)|x=0,\ldots,k\}$. Let
$\rho_{XZ}=\sum_{x=0}^kp_x|x\rangle\langle x|\otimes \rho_x$.
Let $Y$ be a classical random variable which indicates the result of a
measurement on the quantum state with density matrix $\rho_Z=\sum_xp_x\rho_x$. Then
\[H(X:Y)\le\chi({\cal E})=S(X:Z).\]
\end{fact}

We will also need the following lemma.

\begin{lemma}
Let ${\cal E}=\{(p_x,\sigma_x)|x=0,\ldots,k\}$ be an ensemble of
density matrices and let $\sigma=\sum_xp_x\sigma_x$ be the density
matrix of the mixed state of the ensemble. Assume there is an
observable with possible measurement results $x$ and ?, so that for all
$x$ measuring the observable on $\sigma_x$ yields $x$
with probability at least $1-\epsilon$, the result ?~with probability
at most $\epsilon$, and a result $x'\neq x$ with probability 0, then
\[S(\sigma)\ge\sum_xp_xS(\sigma_x)+(1-\epsilon)H(X), \mbox{
  i.e., }\chi({\cal E})\ge (1-\epsilon) H(X).\]
\end{lemma}

\begin{proof} States $x$ of a classical random variable $X$ are coded
as quantum states $\sigma_x$, where
$x$ and $\sigma_x$ have probability $p_x$. The density matrix of the
overall mixed state is $\sigma$ and has von Neumann entropy
$S(\sigma)$. $\sigma$ corresponds to the ``code'' of a random $x$.

According to Holevo's theorem (Fact 4.9) the information on $X$ one can access by
measuring $\sigma$ with result $Y$ is bounded by
$H(X:Y)\le S(\sigma)-\sum_xp_xS(\sigma_x)$. But there is such a
measurement as assumed in the lemma, and with lemma 4.3
$H(X:Y)\ge (1-\epsilon) H(X)$. Thus the lemma follows.\qquad\end{proof}

Not all the relations that are valid in classical information theory
hold in quantum information theory. The following fact states a
notable exception, the so-called Araki-Lieb inequality and one of its consequences,
see \cite{NC00}.

\begin{fact} $S(XY)\ge |S(X)-S(Y)|$.

$S(X:Y|Z)\le 2S(X)$.
\end{fact}

The reason for this behaviour is entanglement.

\begin{lemma} If $\sigma_{XY}$ is separable, then
$S(XY)\ge S(X)$ and
$S(X:Y)\le S(X)$.
\end{lemma}

\subsection{The quantum communication model}

Now we define quantum one-way protocols.

\begin{definition}
In a two player quantum one-way protocol players Alice and Bob each
possess a private set of qubits. Some of the qubits are initialized to
the Boolean inputs of the players, all other qubits are in some fixed
basis state $|0\rangle$.

Alice then performs some quantum operation on her qubits and sends a
set of these qubits to Bob. The latter action changes the possession of
qubits rather than the global state. We can assume that Alice sends
the same number of qubits for all inputs. After Bob has
received the qubits he can perform any quantum operation on the qubits
in his possession and afterwards he announces the result of the
computation. The complexity of a protocol is the number of qubits sent.

In an exact quantum protocol the result has to be correct with
certainty. $Q_E(f)$, is the minimal complexity of an exact quantum
protocol for a function $f$.

In a bounded error protocol the output has to be correct with
 probability $1-\epsilon$ (for $1/2>\epsilon>0$). The bounded error
 quantum one-way communication complexity of a function $f$ is
 $Q_\epsilon(f)$ resp.~$Q(f)=Q_{1/3}(f)$, the minimal complexity of a
 bounded error quantum one-way protocol for $f$.

Quantum Las Vegas protocols are defined regarding acceptance as their
probabilistic counterparts, the notation is $Q_{0,\epsilon}(f)$.

\cite{CB97} considers a different model of quantum
communication: Before the start of the protocol Alice and Bob own a
set of qubits whose state may be
entangled, but must be independent of the inputs. Then as above a
quantum communication protocol is used. We use superscripts $pub$ to
denote the complexity in this model.
\end{definition}

It is possible to simulate the model with entangled qubits by allowing
first an arbitrary finite communication independent of the inputs,
followed by an ordinary protocol.

By measuring distributed EPR-pairs it is possible to simulate classical public randomness.
The technique of superdense coding of \cite{BW92} allows in the model
with prior entanglement to send $n$ bits of classical information with
$\lceil n/2\rceil$ qubits.

\subsection{Quantum circuits and formulae}

Besides quantum Turing machines quantum circuits \cite{D89} are a
 universal model of quantum computation, see \cite{Y93}, and are
 generally easier to handle in descriptions of quantum algorithms. A
 more general model of quantum circuits, in which superoperator gates
 work on density matrices is described in \cite{AKN98}.
 We begin with the basic model.

\begin{definition}
A unitary quantum gate with $k$ inputs and $k$ outputs is specified by a
unitary operator $U:\Com^{2^k}\to\Com^{2^k}$.

A quantum circuit consists of unitary quantum gates with $O(1)$ inputs
and outputs each, plus a set of inputs to the circuits, which are connected
to an acyclic directed graph, in which the inputs are sources. Sources
are labeled by Boolean constants or by input variables.
Edges correspond to qubits, the circuit uses as many
qubits as it has sources. One designated qubit is the output qubit.
A quantum circuit computes a unitary transformation on the source
qubits in the obvious way. In the end the output qubit is measured in
the standard basis.

The size of a quantum circuit is the number of its gates, the depth
is the length of the longest path from an input to the output.

A quantum circuit computes a function with bounded error, if it gives
the right output with probability at least 2/3 for all inputs.

A quantum circuit computes a Boolean function with Monte Carlo error, if it
has bounded error and furthermore never erroneously accepts.

A pair of quantum circuits computes a Boolean function $f$ in the Las Vegas sense,
if the first is a Monte Carlo circuit for $f$, and the second is a
Monte Carlo circuit for $\neg f$.

A quantum circuit computes a function exactly, if it makes no error.
\end{definition}

The definition of Las Vegas circuits is motivated by the fact that we
can easily verify the computation of a pair of Monte Carlo circuits
for $f$ and $\neg f$ as in the classical case, see the proof of  corollary 3.13.

We are interested in restricted types of circuits, namely quantum formulae
\cite{Y93}.

\begin{definition}
A quantum formula is a quantum circuit with the following additional
property: for each source there is at most one path connecting it to
the output. The length or size of a quantum formula is the number of
its sources.
\end{definition}

Apart from the Boolean input variables a quantum formula is allowed to
read Boolean constants only. There is only one final measurement. We
call the model from \cite{Y93} also {\it pure} quantum formulae.

In \cite{AKN98} a more general model of quantum circuits is studied,
in which superoperators work on density matrices.

\begin{definition}
A superoperator gate $g$ of order $(k,l)$ is a trace-preserving,
completely positive map from the density matrices on $k$ qubits to the
density matrices on $l$ qubits.

A quantum superoperator circuit is a directed acyclic graph with inner
vertices marked by superoperator gates with fitting fan-in and
fan-out. The sources are marked with input variables or Boolean
constants. One gate is designated as the output.

A function is computed as follows. In the beginning the sources are
each assigned a density matrix corresponding to the Boolean values
determined by the input or by a constant. The Boolean value 0
corresponds to $|0\rangle\langle 0|$, 1 to $|1\rangle\langle1|$. The
overall state of the qubits involved is the tensor product of these
density matrices.

Then the gates are applied in an arbitrary topological order. Applying
a gate means applying the superoperator composed of the gates'
superoperator on the chosen qubits for the gate and the identity
superoperator on the remaining qubits.

In the end the state of the output qubit is supposed to be a classical
probability distribution on
$|0\rangle$ and $|1\rangle$.
\end{definition}

The following fact from \cite{AKN98} allows to apply gates in an
arbitrary topological ordering.

\begin{fact}
Let $C$ be a quantum superoperator circuit, $C_1$ and $C_2$ be two
sets of gates working on different sets of qubits. Then for all
density matrices $\rho$ on the qubits in the circuit the result of
$C_1$ applied to the result of $C_2$ on $\rho$ is the same as the result of
$C_2$ applied to the result of $C_1$ on $\rho$.

Let two arbitrary topological orderings of the gates in a quantum
superoperator circuit be given. The result of applying the gates in
one ordering is the same as the result of applying the gates in the
other ordering for any input density matrix.
\end{fact}

One more aspect is interesting in the definition of quantum
formulae: we want to allow quantum formulae to access multiple
read random inputs, just as fair probabilistic formulae. This
makes it possible to simulate the latter model.  Instead of random
variables we allow the quantum formulae to read an arbitrary
nonentangled state. A pure state on $k$ qubits is called
nonentangled, if it is the tensor product of $k$ states on 1 qubit
each. A mixed state is nonentangled, if it can be expressed as a
probabilistic ensemble of nonentangled pure states. Note that a
classical random variable read $k$ times can be modelled as
$|1^k\rangle$ with probability 1/2 and $|0^k\rangle$ with
probability 1/2.

We restrict our definition to gates with fan-in 2, the set of
quantum gates with fan-in 2 is known to be universal \cite{BBC95}.

\begin{definition} A generalized quantum formula is a quantum
  superoperator circuit with fan-out 1/fan-in 2
  gates together with a fixed nonentangled mixed state. The sources of the circuit are either labeled by input
  variables, or may access a qubit of the state. Each qubit of this state may be accessed only by one gate.
\end{definition}

As proved in \cite{AKN98} the Kraus representation theorem (Fact 4.7) implies that quantum superoperator circuits with constant
fan-in are asymptotically as efficient as quantum circuits with
constant fan-in. The same holds for quantum formulae. The
essential difference between pure and generalized quantum formulae
is the availability of multiple read random bits.

\subsection{Programmable quantum gates}

For simulations of quantum mechanical formulae by communication
protocols we will need a program\-mable quantum gate. Such a gate
allows Alice to communicate a unitary operation as a program
stored in some qubits to Bob, who then applies this operation to
some of his qubits.

Formally we have to look for a unitary operator $G$ with
\[G(|d\rangle\otimes |P_U\rangle)=U(|d\rangle)\otimes |P'_U\rangle.\]
Here $|P_U\rangle$ is the "code"\ of a unitary operator $U$, and
$|P'_U\rangle$ the some leftover of the code.

The bad news is that such a programmable gate does not exist, as proved
in \cite{NC97}. Note that in the classical case such gates
are easy to construct.

\begin{fact}
If $N$ different unitary operators (pairwise different by more than a
global phase) can be implemented by a programmable quantum gate, then
the gate needs a program of length $\log N$.\end{fact}

Since there are infinitely many unitary operators on just one qubit
there is no programmable qubit with finite program length implementing
them all. The proof uses that the gate works deterministically, and
actually a probabilistic solution to the problem exists.

We now sketch a construction of  Nielsen and Chuang \cite{NC97}.
For the sake of simplicity we just describe the construction for
unitary operations on one qubit.

The program of a unitary operator $U$ is
\[|P_U\rangle=\frac{1}{\sqrt{2}}(|0\rangle U|0\rangle + |1\rangle U|1\rangle).\]

The gate receives as input $|d\rangle\otimes |P_U\rangle$. The gate
then measures the first and second qubit in the basis
$\{|\Phi^+\rangle,|\Phi^-\rangle,|\Psi^+\rangle,|\Psi^-\rangle\}$.
Then the third qubit is used as a result.

For a state $|d\rangle = a|0\rangle+b|1\rangle$ the input to the gate is
\begin{eqnarray*}
& &[a|0\rangle+b|1\rangle]\frac{|0\rangle U|0\rangle + |1\rangle U|1\rangle}{\sqrt{2}}\\
&=&\frac{1}{2}\left[|\Phi^+\rangle (a U|0\rangle+b U|1\rangle) +
|\Phi^-\rangle (a U|0\rangle-b U|1\rangle)\right.\\
& &\left. +|\Psi^+\rangle (a U|1\rangle+b U|0\rangle) +
|\Psi^-\rangle (a U|1\rangle-b U|0\rangle)\right].\end{eqnarray*}

Thus the measurement produces the correct state with probability
1/4 and moreover the result of the measurement indicates whether
the computation was done correctly. Also, given this measurement
result we know exactly which unitary "error" operation has been
applied before the desired operation. We now state Nielsen and
Chuang's result.

\begin{fact} There is a probabilistic programmable quantum gate with
  $m$ input qubits for the state plus $2m$ input qubits for the
  program, which implements every unitary operation on $m$ qubits, and
  succeeds with probability $1/2^{2m}$. The result of a measurement
  done by the gate indicates whether the computation was done
  correctly, and which unitary error operation has been performed.
\end{fact}

\section{One-way communication complexity: the nondeterministic and
  the quantum case}

\subsection{A lower bound for  limited nondeterminism}

In this section we investigate nondeterministic one-way communication
with a limited number of nondeterministic bits. Analogous problems for
many round communication complexity have been addressed in
\cite{HS96}, but in this section we again consider asymmetric
problems, for which the one-way restriction is essential.

It is easy to see that if player Bob has $m$ input bits then $m$
nondeterministic bits are the maximum player Alice needs. Since the
nondeterministic communication complexity without any
limitation on the number of available nondeterministic bits is at most
$m$, Alice can just guess the communication and send it to
Bob in case it is correct with respect to her input and leads to
acceptance. Bob can then check the same for his input.
Thus an optimal protocol can be simulated.

For the application to lower bounds on formula size we are again
interested in functions with an asymmetric input partition, i.e.,
Alice receives much more inputs than Bob. For nontrivial results thus
the number of nondeterministic bits must be smaller than the number of
Bob's inputs.

A second observation is that using $s$ nondeterministic bits can
reduce the communication complexity from the deterministic one-way
communication complexity $d$ to $d/2^s$ in the best case. If $s$ is
sublogarithmic, strong lower bounds follow already from the
deterministic lower bounds, e.g.~$N_{\epsilon\log n}(\neg EQ)\ge
n^{1-\epsilon}$, while $N_{\log n}(\neg EQ)=O(\log n)$. On the other
hand:

\begin{lemma}
\[N_s(f)=c \Rightarrow N_c\le c.\]
\end{lemma}

\begin{proof} In a protocol with communication $c$ at most $2^c$
 different messages can be sent (for all inputs). To guess such a
 message $c$ nondeterministic bits are sufficient.\qquad\end{proof}

It is not sensible to guess more than to communicate. We are
interested in determining how large the difference between
nondeterministic one-way communication complexity with $s$
nondeterministic bits and unrestricted nondeterministic
communication complexity may be. Therefore we consider the maximal
such gap as a function $G$.

\begin{corollary}
Let $f:\{0,1\}^n\times\{0,1\}^m\to\{0,1\}$  be a Boolean function and
$G:\N\to\N$ a monotone increasing function with

$N(f)=c$ and $N_s(f)=G(c)$ for some $s$.

Then $N_{G^{-1}(n)}(f)\le c$ and hence $s\le G^{-1}(n)$, where
$G^{-1}(x)=\min\{y|G(y)\ge x\}$.
\end{corollary}

\begin{proof} $G(c)\le n$ and hence $c\le G^{-1}(n)$.\qquad\end{proof}

The range of values of $s$, for which a gap $G$ between $N(f)$ and
$N_s(f)$ is possible is thus limited. If e.g.~an exponential
difference $G(x)=2^x$ holds, then
$s\le\log n$. If $G(x)=r\cdot x$, then $s\le n/r$.

We now show a gap between nondeterministic one-way communication
complexity with $s$ nondeterministic bits and unlimited
nondeterministic communication complexity. First we define the
family of functions exhibiting this gap.

\begin{definition}
Let $D_{n,s}$ be the following  Boolean function for $1\le s\le n$:

\begin{eqnarray*}D_{n,s}(x_1,\ldots,x_n;x_{n+1})=1\iff \forall i: x_i\in {\cal
P}(n^3,s) \\ \wedge\exists i:
|\{j|j \neq i;x_i\cap x_j\neq\emptyset\}|\ge s.\end{eqnarray*}
\end{definition}

Note that the function has $\Theta(sn\log n)$ input bits in a standard
encoding.
We consider the partition of inputs in which Bob receives the set
$x_{n+1}$ and Alice all other sets. The upper bounds in the following
lemma are trivial, since Bob only receives $O(s\log n)$ input bits.

\begin{lemma}
\[N_{O(s\log n)}  (D_{n,s})=O(s\log n).\]

\[D^{B}(D_{n,s})=O(s\log n).\]
\end{lemma}

The lower bound we present now results in a near optimal difference
between nondeterministic (one-way) communication and limited
nondeterministic one-way communication.
Limited nondeterministic one-way communication has also been studied
subsequently to this work in \cite{HrSa00}. There a tradeoff between
the consumption of nondeterministic bits and the one-way
communication is demonstrated (i.e., with more nondeterminism the
communication gradually decreases). Here we describe a fundamentally
different phenomenon of a threshold type: nondeterministic bits do
not help much, until a certain amount of them is available, when
quite quickly the optimal complexity is attained. For more
results of this type see \cite{Kl98}.

\begin{theorem} There is a constant
$\epsilon>0$, so that for $s\le n$
\[N_{\epsilon s}  (D_{n,s})=\Omega(ns\log n).\]
\end{theorem}

\begin{proof} We have to show that all nondeterministic one-way
protocols  computing $D_{n,s}$ with $\epsilon s$ nondeterministic
bits need much communication.

A nondeterministic one-way protocol with $\epsilon s$ nondeterministic
bits and communication $c$ induces a cover of the communication matrix
with $2^{\epsilon s}$ Boolean matrices having the following properties:
each 1-entry of the communication matrix is a 1-entry in at least one
of the Boolean matrices, no 0-entry of the communication matrix is a
1-entry in any of the Boolean matrices, furthermore the set of rows
appearing in those matrices has size at most $2^c$. This set of
matrices is obtained by fixing the nondeterministic bits and taking
the communication matrices of the resulting deterministic protocols.
We show the lower bound from the property that {\it each} of the
Boolean matrices covering the communication matrices uses at most
$2^c$ different rows. Thus the lower bound actually even holds for
protocols with limited, but public nondeterminism.

We first construct a submatrix of the communication matrix with some
useful properties, and then show the theorem for this
``easier'' problem.

Partition the universe $\{1,\ldots,n^3\}$ in $n$ disjoint sets
$U_1,\ldots,U_n$ with $|U_i|=n^2=m$.
Then choose vectors of $n$ size $s$ subsets of the universe, so that the $i$th
subset is from $U_i$. Thus the $n$ subsets of a vector are pairwise
disjoint. Now the protocol has to determine, whether the set of Bob
intersects nontrivially with $s$ sets of Alice.

We restrict the set of inputs further. There are $m\choose s$
subsets of $U_i$ having size $s$. We choose a set of such subsets so
that each pair of them have no more than $s/2$ common elements.
To do so we start with any subset and remove all subsets in
``distance'' at most $s/2$. This continues as long as possible. We get
a set of subsets of $U_i$, whose elements have pairwise distance at
least $s/2$. In every step at most ${s\choose s/2}{m\choose s/2}$
subsets are removed, thus we get at least
\begin{eqnarray}\frac{{m\choose s}}{{s\choose s/2}{m\choose
  s/2}}\ge\left(\frac{m}{s}\right)^{s/2}/2^{3s/2}\end{eqnarray}
sets.

As described we draw Alice's inputs as vectors of sets, where the set
at position $i$ is drawn from the set of subsets of
$U_i$ we have just constructed. These inputs are identified with the
rows of the submatrix of the communication matrix.
The columns of the submatrix are restricted to elements of
$U_1\cup\{\top\}\times\cdots\times U_n\cup\{\top\}$, for which $s$ positions are
occupied, i.e., $n-s$ positions carry the extra symbol
$\top$ which stands for ``no element''.
Call the constructed submatrix $M$.

Now assume there is a protocol computing the restricted
problem. Fixing the nondeterministic bits induces a deterministic
protocol and a matrix $M'$, which covers at least
$1/2^r$ of the ones of $M$, where $r=\epsilon s$.
We now show that such a matrix must have many different rows, which
corresponds to large communication.

Each row of $M$ corresponds to a vector of $n$ sets.
A position $i$ is called a {\it difference position} for a pair of
such vectors, if they have different sets at position $i$. According
to our construction these sets have no more than $s/2$ elements in common.

We say a set of rows has $k$ difference positions, if there are $k$
positions $i_1,\ldots,i_k$, so that for each $i_l$ there are two rows
in the set for which $i_l$ is a difference position.

We now show that each row of $M'$ containing ``many'' ones does not
``fit'' on many rows of $M$, i.e., contains ones these do not
have. Since $M'$ has one-sided error only, the rows of $M'$ are either
sparse or cover only few rows of $M$. Observe that each row of $M$ has
exactly ${n\choose s} s^{s}$ ones.

\begin{lemma}
Let $z$ be a row of $M'$, appearing several times in $M'$. The rows of
$M$, in whose place in $M$ the row $z$ appears in $M'$, may have
$\delta n$ difference positions. Then $z$ contains at most $2{n\choose
  s} s^{s}/2^{\delta s/6}$ ones.
\end{lemma}

\begin{proof}
Several rows of $M$ having $\delta n$ difference positions are given,
and the ones of $z$ occur in all of these rows. Let $C$ be the set of
${n\choose s}s^s$ columns/sets being the ones in the first such
row. All other columns are forbidden and may not be ones in $z$.

A column in $C$ if chosen randomly by choosing $s$ out of $n$
positions and then one of $s$ elements for each position. Let $k=\delta s$.
We have to show an upper bound on the number of ones in $z$, and we
analyze this number as the probability of getting a one when choosing
a column in $C$. The probability of getting a one is at most the
probability that the chosen positions have a nontrivial intersection
with less than $k/2$ sets $U_i$ at difference positions $i$ (event $E$) plus the
probability of getting a one under the condition of event $\overline{E}$,
following the general formula $Prob(A)\le Prob(A|E)+Prob(\overline{E})$.

We first count the columns in $C$, which have a nontrivial
intersection with at most $k/2$ of the sets $U_i$ at difference
positions $i$. Consider the slightly different experiment in which $s$ times
independently one of $n$ positions is chosen, hence positions may
be chosen more than one time. Now expected $\delta s=k$ difference
positions are chosen. Applying Chernov's inequality yields that with
probability at most
\[e^{-\frac{1}{4\cdot 2}\cdot k}\le 2^{-\delta s/6}\] at most $k/2$
difference positions occur. When choosing a random column in $C$
instead, this probability is even smaller, since now positions are
chosen without repetitions. Thus the columns in $C$, which ``hit'' less
than $k/2$ difference positions, contribute at most $2^{-\delta
  s/6}{n\choose s}s^s$ ones to $z$.

Now consider the columns/sets in $C$, which intersect at least $k/2$
of the $U_i$ at difference positions $i$. Such a column/set fits on
all the rows, if the element at each position not bearing a $\top$
lies in the intersection of all sets in the rows at position $i$. At
each difference position there are two rows, which hold different sets
at that position, and those sets have distance $s/2$.

Fix an arbitrary set of positions such that at least $k/2$ difference positions
are included. The next step of choosing a column in $C$ consists of
choosing one of $s$ elements for each position. But if a position is a
difference position, then at most $s/2$ elements satisfy the condition
of lying in the sets held by all the rows at that position. Thus the probability of
fitting on all the rows is at most $2^{-k/2}$, and at most ${n\choose
s}s^s/2^{k/2}$ such columns can be a one in $z$.

Overall only a fraction of $2^{-\delta s/6+1}$ of all columns in $C$
can be ones in $z$.\qquad\end{proof}

At least one half of all ones in $M'$ lie in rows containing at least
$\ge{n\choose s}s^s/2^{r+1}$ ones. Lemma 5.6 tells us that such a row
fits only on a set of rows of $M$ having no more than $\delta n$
difference positions, where $r+1=\delta s/6-1$. Hence such a row can cover
at most all the ones in
${m \choose s}^{\delta n}$ rows of $M$, and therefore only ${m \choose s}^{\delta n}{n
\choose s} s^s$ ones.

According to (5.1) at least
$(m/s)^{sn/2}{n\choose s} s^s/(2^{3sn/2}2^{r+1})$ ones are covered by
such rows, hence
\begin{eqnarray*}
&&\frac{(m/s)^{sn/2}{n\choose s}s^s}{{m \choose s}^{\delta n}{n
\choose s} s^s2^{3sn/2}2^{r+1}}\\
&&\ge\frac{(m/s)^{sn/2}}{(em/s)^{6\epsilon sn+12n}2^{3sn/2}2^{\epsilon s+1}}\\
&&=2^{\Omega(sn\log n)}\end{eqnarray*}
rows are necessary (for $\epsilon=1/20$ and $n\ge s\ge 400$).\qquad\end{proof}

\subsection{Quantum one-way communication}

Our first goal in this section is to prove that the VC-dimension lower
bound for randomized one-way protocols (Fact 2.9) can be extended to the quantum
case. To achieve this we first prove a linear lower bound on the
bounded error quantum communication complexity of the index
function $IX_n$, and then describe a reduction from the index function
$IX_d$ to any function with VC-dimension $d$, thus transferring the
lower bound. It is easy to see that $VC(IX_n)=n$, and thus the
bounded error  probabilistic one-way communication complexity is large
for that function.

The problem of {\it random access quantum coding} has been considered
in \cite{ANTV99} and \cite{Na99}. In a
$n,m,\epsilon$-random access quantum code all Boolean
$n$-bit words $x$ have to be mapped to states of $m$ qubits each, so
that for
$i=1,\ldots,n$ there is an observable, so that measuring the
quantum code with that observable yields the bit
$x_i$ with probability $1-\epsilon$. The quantum code is allowed to be
a mixed state. Nayak
\cite{Na99} has shown
\begin{fact}
For every $n,m,\epsilon$-random access quantum coding $m\ge(1-H(\epsilon))n$.
\end{fact}

It is easy to see that the problem of random access quantum coding is
equivalent to the construction of a quantum one-way protocol for the
index function. If there is such a protocol, then the messages can
serve as mixed state codes, and if there is such a code the codewords
can be used as messages.
We can thus deduce a lower bound for $IX_n$ in the model of one-way
quantum communication complexity without prior entanglement.

We now give a proof, that can also be adapted to the case of allowed prior entanglement.

\begin{theorem}
$Q_\epsilon (IX_n)\ge (1-H(\epsilon))n$.

$Q^{pub}_\epsilon (IX_n)\ge (1-H(\epsilon))n/2$.
\end{theorem}

\begin{proof} Let $M$ be the register containing the message sent by
Alice, and let $X$ be a register holding a uniformly random input to
Alice. Then $\sigma_{XM}$ denotes the state of Alice's qubits directly
before the message is sent. $\sigma_M$ is the state of a random
message. Now every bit is decodable with
probability $1-\epsilon$ and thus
$S(X_i:M)\ge 1-H(\epsilon)$ for all $i$. To see this consider $S(X_i:M)$
as the Holevo information of the following ensemble:
\[\sigma_{i,0}=\sum_{x:x_i=0}\frac{1}{2^{n-1}}\sigma_{M}^x\] with
probability 1/2 and
\[\sigma_{i,1}=\sum_{x:x_i=1}\frac{1}{2^{n-1}}\sigma_{M}^x\] with
probability 1/2, where $\sigma_{M}^x$ is the density matrix of the
message on input $x$. The information obtainable on $x_i$ by measuring
$\sigma_M$ must be at
$1-H(\epsilon)$ due to Fano's inequality Fact 4.2, and thus the Holevo
information of the ensemble is at least $1-H(\epsilon)$, hence
$S(X_i:M)\ge 1-H(\epsilon)$.

But then $S(X:M)\ge (1-H(\epsilon)n$ (since all $X_i$
are mutually independent). $S(X:M)\le S(M)$ using lemma 4.12, since
$X$ and $M$ are not entangled. Thus the number of qubits in $M$ is at
least $(1-H(\epsilon))n$.

Now we analyze the complexity of
$IX_n$ in the one-way communication model with entanglement.

The density matrix of the state induced by a uniformly random input on
$X$, the message $M$, and the qubits $E_A,E_B$ containing the prior
entanglement in the possession of Alice and Bob, is
$\sigma_{XME_AE_B}$. Here $E_A$ contains those qubits of the entangled
state Alice keeps, note that some of the entangled qubits will usually
belong to $M$.
Tracing out $X$ and $E_A$ we receive a state $\sigma_{ME_B}$, which is
accessible to Bob. Now every bit of the string in $X$ is decodable, thus
$S(X_i:ME_B)\ge 1-H(\epsilon)$ for all $i$ as before.
But then also $S(X:ME_B)\ge(1-H(\epsilon)n$, since all the $X_i$ are
mutually independent.

$S(X:ME_B)=S(X:E_B)+S(X:M|E_B)\le 2S(M)$ by an application of the
Araki-Lieb inequality, see Fact 4.11. Note that $S(X:E_B)=0$. So the number
of qubits in $M$ must be at least $(1-H(\epsilon))n/2$.\qquad\end{proof}

Note that the lower bound shows that 2-round deterministic
communication complexity can be exponentially smaller than one-way
quantum communication complexity. For a more general quantum
communication round-hierarchy see \cite{KNTZ01}.

\begin{theorem}
For all functions $f:$
$Q_\epsilon(f)\ge(1-H(\epsilon))VC(f)$ and

$Q^{pub}_\epsilon(f)\ge(1-H(\epsilon))VC(f)/2$.
\end{theorem}

\begin{proof} We now describe a reduction from the index function to $f$. Assume
$VC(f)=d$, i.e., there is a set $S=\{s_1,\ldots,s_d\}$ of inputs for
Bob, which is shattered by the set of functions $f(x,.)$. The reduction then goes
from $IX_d$ to $f$.

For each $R\subseteq S$ let $c_R$ be the incidence vector of $R$
(having length $d$).
$c_R$ is a possible input for Alice when computing the index function $IX_d$.
For each $R$ choose some $x_R$, which separates this subset from the
rest of $S$, i.e., so that $f(x_R,y)=1$ for all $y\in R$ and
$f(x_R,y)=0$ for all $y\in S-R$.

Assume a protocol for $f$ is given. To compute the index function the
players do the following. Alice maps $c_R$ to $x_R$.
Bob's inputs $i$ are mapped to the $s_i$. Then $f(x_R,s_i)=1\iff
s_i\in R\iff c_R(i)=1$.

In this manner a quantum protocol for $f$ must implicitly compute
$IX_d$. According to Theorem 5.8 the lower bounds follow.\qquad\end{proof}

Application of the previous theorem gives us lower bounds for the
disjointness problem in the model of quantum one-way communication
complexity. Lower bounds of the order $\Omega(n^{1/k})$ for constant
$k$ in $k$-round protocols are given in \cite{KNTZ01}.

\begin{corollary}
$Q_\epsilon (DISJ_n)\ge (1-H(\epsilon))n$.

$Q_\epsilon^{pub}(DISJ_n)\ge (1-H(\epsilon))n/2$.
\end{corollary}

The first result has independently been obtained in  \cite{BW01}.
Note that the obtained lower bound method is not tight in general.
There are functions for which an unbounded gap exists between the
VC-dimension and the quantum one-way communication complexity
\cite{Kl00b}.

Now we turn to the exact and Las Vegas quantum one-way communication
complexity. For classical one-way protocols it is known that Las Vegas
communication complexity is at most a factor 1/2 better than
deterministic communication for total functions, see Fact 2.10.

\begin{theorem}
For all total functions $f$:

$Q_E (f)=D(f)$,

$Q_{0,\epsilon} (f)\ge (1-\epsilon) D(f)$.
\end{theorem}

\begin{proof}
Let $row(f)$ be the number of different rows in the communication
matrix of $f(x,y)$. According to Fact 2.7 $D(f)=\lceil\log
row(f)\rceil$. We assume in the following that the communication
matrix consists of pairwise different rows only.

We will show that any Las Vegas one-way protocol which gives up with
probability at most $\epsilon\ge 0$ for some function $f$ having
$row(f)=R$, must use messages with von Neumann entropy at least
$(1-\epsilon)\log R$, when started on a uniformly random input. Inputs
for Alice are identified with rows of the communication matrix.
We then conclude that the Hilbert space of the messages must have
dimension  at least $R^{1-\epsilon}$ and hence at least
$(1-\epsilon)\log R$ qubits have to be sent. This gives us the second
lower bound of the theorem. The upper bound of the first statement is
trivial, the lower bound of the first statement follows by taking $\epsilon=0$.

We now describe a process, in which rows of the communication matrix
are chosen randomly bit per bit. Let $p$ be the probability of having
a 0 in column 1 (i.e., the number of 0s in column 1 divided by the number of
rows). Then a 0 is chosen with probability $p$, a 1 with probability
$1-p$. Afterwards the set of rows is partitioned into the set $I_0$ of
rows starting with a 0, and the set $I_1$ of rows starting with a
1. When $x_1=b$ is chosen, the process continues with $I_b$ and the
next column.

Let $\rho_y$ be the density matrix of the following mixed state: the
(possibly mixed) message corresponding to a row starting with $y$ is chosen
uniformly over all such rows.

The probability, that a 0 is chosen after $y$ is called
$p_y$, and the number of different rows beginning with $y$ is called $row_y$.

We want to show via induction that $S(\rho_y)\ge(1-\epsilon)\log row_y$.
Surely $S(\rho_y)\ge 0$ for all $y$.

Recall that Bob can determine the function value for an arbitrary
column with the correctness guarantee of the protocol.

Then with lemma 4.10 $S(\rho_y)\ge p_y
S(\rho_{y0})+(1-p_y)S(\rho_{y1})+(1-\epsilon)H(p_y)$, and via induction
\begin{eqnarray*}
&S(\rho_y)&\ge p_y((1-\epsilon)\log row_{y0})\\& &+(1-p_y)((1-\epsilon)\log
row_{y1})+(1-\epsilon)H(p_y)\\
& &=(1-\epsilon)[p_y\log(p_yrow_y)\\& &+(1-p_y)\log((1-p_y)row_y)+H(p_y)]\\
& &=(1-\epsilon)\log row_y.\end{eqnarray*}
We conclude that $S(\rho)\ge(1-\epsilon)\log row(f)$ for the density
matrix $\rho$ of a message to a uniformly random row. Hence the lower bound on the
number of qubits holds.
\qquad\end{proof}

We now again consider the model with prior entanglement.

\begin{theorem} For all total functions $f$:

$Q^{pub}_E(f)=\lceil D(f)/2\rceil,$

$Q^{pub}_{0,\epsilon}(f)\ge D(f)(1-\epsilon)/2$.
\end{theorem}

The upper bound follows from superdense coding \cite{BW92}.
Instead of the lower bounds of the theorem we prove a stronger
statement. We consider an extended model of quantum one-way
communication, that will be useful later.

In a {\it nonstandard} one-way quantum protocol Alice and Bob are allowed to
communicate in arbitrarily many rounds, i.e., they can exchange
many messages. But Bob is not allowed to send Alice a message, so that
the von Neumann information between the input of Alice plus the accessible
qubits of Alice and Bob's input is larger than 0. The communication
complexity of a protocol is the number of qubits sent by Alice in the
worst case. The model is at least as powerful as the model with prior
entanglement, since Bob may e.g.~generate some EPR-pairs, send one qubit of
each pair to Alice, then Alice may send a message as in a protocol
with prior entanglement.

\begin{lemma}
For all functions $f$ a nonstandard quantum one-way protocol with
bounded error must communicate at least $(1-H(\epsilon))VC(f)/2$
qubits from Alice to Bob.

For all total functions $f$ a nonstandard quantum one-way protocol
\begin{remunerate}
\item with exact acceptance must communicate at least $\lceil
  D(f)/2\rceil$ qubits from Alice to Bob.
\item with Las Vegas acceptance and success probability $1-\epsilon$
  must communicate at least
  $ (1-\epsilon)D(f)/2$ qubits from Alice to Bob.
\end{remunerate}
\end{lemma}

\begin{proof}
In this proof we always call the qubits available to Alice $P$, and
the qubits available to Bob $Q$, for simplicity disregarding that
these registers change during the course of the protocol. We assume that the inputs
are in registers $X,Y$ and are never erased or changed in the
protocol. Furthermore we assume that for all fixed values $x,y$ of the
inputs the remaining global state is pure.

For the first statement it is again sufficient to investigate the
complexity of the index function.

Let $\sigma_{XYPQ}$ be the state for random inputs in $X,Y$ for Alice
and Bob, with qubits $P$ and $Q$ in the possession of Alice and Bob.
Since Bob determines the result, it must
be true that in the end of the protocol
$S(X_Y:YQ)\ge 1-H(\epsilon)$, since the value $X_Y$ can be determined
from Bob's qubits with probability $1-\epsilon$. It is always true in
the protocol that $S(XP:Y)=0$. Let $\rho_P^{X=x,Y=y}$ be the density
matrix of $P$ for fixed inputs $X=x$ and $Y=y$. Then we
have that for all $x,y,y'$: $\rho_P^{X=x,Y=y}=\rho_P^{X=x,Y=y'}$.

$\rho_{PQ}^{X=x,Y=y}$ purifies $\rho_P^{X=x,Y=y}$. Then the
following fact from \cite{M97} and \cite{LC98} tells us that all $y$
and corresponding states of $Q$ are ``equivalent'' from the perspective of Alice.

\begin{fact}
Assume $|\phi_1\rangle$ and $|\phi_2\rangle$ are pure states in a
Hilbert space
$H\otimes K$, so that $Tr_K |\phi_1\rangle\langle\phi_1|=Tr_K
|\phi_2\rangle\langle\phi_2|$.

Then there is a unitary transformation $U$ acting on $K$, so that
$I\otimes U |\phi_1\rangle=|\phi_2\rangle$ (for the identity operator $I$ on $H$).
\end{fact}

Thus there is a local unitary transformation applicable by Bob alone, so
that $\rho_{PQ}^{X=x,Y=y}$ can be changed to $\rho_{PQ}^{X=x,Y=y'}$
Hence for all  $i$ we have
$S(QY:X_i)\ge1-H(\epsilon)$, and thus $S(X:QY)\ge (1-H(\epsilon))n$.

In the beginning $S(X:QY)=0$. Then the protocol proceeds w.l.o.g.~so
that each player applies a unitary transformation on his qubits and
then sends a qubit to the other player. Since
the information cannot increase by local operations, it is sufficient
to analyze what happens if qubits are sent. When
Bob sends a qubit to Alice $S(X:QY)$ is not increased. When Alice sends
a qubit to Bob, then $Q$ is augmented by a qubit $M$, and $S(X:QMY)\le
S(X:QY)+S(XQY:M)\le S(X:QY)+2S(M)\le S(X:QY)+2$ due to Fact 4.11. Thus the
information can increase only when Alice sends a qubit and always by
at most 2. The lower bound follows.

Now we turn to the second part. We consider the same situation as in
the proof of Theorem 5.11.
Let  $\sigma^{rc}_P$ denote the density matrix of the qubits $P$ in
Alice's possession under the condition that the input row is $r$ and
the input column is $c$. Clearly $\sigma^{rc}_{PQ}$ (containing also
Bob's qubits) is a purification
of $\sigma^{rc}_P$. Again $\sigma^{rc}_P=\sigma_P^{rc'}$ for all
$r,c,c'$, and according to Fact 5.14 for all $c$ and all corresponding
states of $Q$, it is true that Bob can switch locally between
them. Hence it is possible for Bob to compute the function for an
arbitrary column.

The probability of choosing a 0 after a prefix $y$ of a row is again called $p_y$, and
the number of different rows beginning with $y$ is called
$row_y$. $\rho_y$ contains the state of Bob's qubits at the end of the
protocol if a random row starting with $y$ is chosen uniformly (and
some fixed column $c$ is chosen).
Surely $S(\rho_y)\ge 0$ for all $y$. Since Bob can change his column
(and the corresponding state of $Q$) by a local unitary
transformation, he is able to compute the function for an arbitrary column, always
with the success probability of the protocol, at the end. With lemma
4.10 $S(\rho_y)\ge p_y
S(\rho_{y0})+(1-p_y)S(\rho_{y1})+(1-\epsilon)H(p_y)$.

At the end of the protocol thus $S(\sigma_Q^c)=S(\rho)\ge(1-\epsilon)\log
row(f)+\sum_r\frac{1}{row(f)}S(\sigma_Q^{rc})$ for all $c$.
Thus the Holevo information of the ensemble, in which
$\rho_r=\sigma_Q^{rc}$ is
chosen with probability $1/row(f)$ is at least $(1-\epsilon)\log
row(f)$. Let $\sigma_{RPQ}$ be the density matrix of rows, qubits of
Alice and Bob. It follows that $S(R:Q)\ge(1-\epsilon)\log row(f)$ and
as before at least half that many qubits have to be sent from Alice
to Bob.
\qquad\end{proof}

\section{More lower bounds on formula size}

\subsection{Nondeterminism and formula size}

Let us first mention that any nondeterministic circuit can easily be
transformed into a nondeterministic formula without increasing size by
more than a constant factor. To do so one simply guesses the values  of
all gates and then verifies that all guesses are correct and that the
circuit accepts. This is a big AND over test involving $O(1)$
variables, which can be implemented by a CNF each. Hence lower bounds for
nondeterministic formulae are very hard to prove, since even nonlinear
lower bounds for the size of deterministic circuits computing some
explicit functions are unknown. We now show that formulae with limited
nondeterminism are more accessible.
We start by introducing a variant of the Ne\v ciporuk method, this
time with nondeterministic communication:

\begin{definition}
Let $f$ be a Boolean function with $n$ input variables and
$y_1\ldots y_k$ be a partition of the inputs in $k$ blocks.

Player Bob receives the inputs in $y_i$ and player Alice receives all
other inputs.
The nondeterministic one-way communication complexity of with $s$
nondeterministic bits
of $f$ under this input partition is called $N_s (f_i)$.
Define the $s$-nondeterministic Ne\v ciporuk function as
$1/4\sum_{i=1}^{k} N_s (f_i)$.
\end{definition}

\begin{lemma}
The $s$-nondeterministic Ne\v ci\-po\-ruk function is a lower bound
for the length of nondeterministic Boolean formulae with $s$
nondeterministic bits.
\end{lemma}

 The proof is analogous to the proof of Theorem 3.5. Again protocols
 simulate the formula in $k$ communication games. This time Alice
 fixes the nondeterministic bits by herself, and no probability
 distribution on formulae is present.

We will apply the above methodology to the following language.

\begin{definition}
Let $AD_{n,s}$ denote the following language (for $1\le s\le n$):
\begin{eqnarray*}AD_{n,s}=\{(x_1,\ldots,x_{n+1})|\forall i: x_i\in {\cal
P}(n^3,s),\\
x_i \mbox{ is written in sorted order}
\\\wedge\exists i:
|\{j|j \neq i;x_i\cap x_j\neq\emptyset\}|\ge s\}.\end{eqnarray*}
\end{definition}

\begin{theorem}
Every nondeterministic formula with $s$ nondeterministic bits
for $AD_{n,20s}$ has length at least
$\Omega(n^2s\log n)$.

$AD_{n,s}$ can be computed by a nondeterministic formula of length
$O(ns^2\log n)$, which uses $O(s\log n)$ nondeterministic bits (for $s\ge\log n$).
\end{theorem}

\begin{proof} For the lower bound we use the methodology we have just
described. We consider the $n+1$ partitions of the inputs, in which
Bob receives the set $x_i$ and Alice all other sets.
The function they have to compute now is the function $D_{n,s}$ from
definition 5.3. In Theorem 5.5 a lower bound of $\Omega(ns\log n)$ is
shown for this problem, hence the length of the formula is
$\Omega(n\cdot ns\log n)$.

For the upper bound we proceed as follows: the formula guesses (in
binary) a number $i$ with $1\le i\le n+1$ and pairs
$(j_1,w_1),\ldots,(j_s,w_s)$, where $1\le j_k\le n+1$ and $1\le w_k\le
n^3$ for all $k=1,\ldots,s$. The number $i$ indicates a set, and the
pairs are witnesses that set $i$ and set $j_k$ intersect on element $w_k$.

The formula does the following tests. First there is a test, whether
all sets consist of $s$ sorted elements. For this $ns$ comparisons of
the form $x_i^j<x_i^{j+1}$ suffice, which can be realized with
$O(\log^2n)$ gates each. Since $s\ge\log n$ overall $O(ns^2\log n)$
gates are enough.

The next test is, whether $j_1<\cdots<j_s$. This makes sure that
witnesses for $s$ different sets have been guessed. Also $i\neq j_k$
for all $k$ must be tested.

Then the formula tests, whether for all $1\le l\le n+1$ the following
holds: if  $l=i$, then all guessed elements are in $x_l$; if $1\le
l\le n+1$ and $1\le k\le s$ the formula also tests, whether $l=j_k$
implies, that $w_k\in x_l$.

All these test can be done simultaneously by a formula of length $O(ns^2\log n)$.\qquad\end{proof}

For $0<\epsilon\le 1/2$ let
$s=n^\frac{\epsilon}{1-\epsilon}$, then the lower bound for limited
nondeterministic formulae is $\Omega(N^{2-\epsilon}/\log^{1-\epsilon} N)$ with
$N^{\epsilon}/\log^\epsilon N$ nondeterministic bits allowed. $O(N^{\epsilon}\log^{1-\epsilon}
N)$ nondeterministic bits suffice to construct a formulae having length
$O(N^{1+\epsilon}/\log^{\epsilon} N)$. Hence the threshold for
constructing an efficient formula is polynomially large, allowing an
exponential number of computations on each input.

\subsection{Quantum formulae}

Now we derive lower bound for generalized quantum formulae. In
\cite{RV01} pure quantum formulae are considered (recall these are
quantum formulae which may not access multiply readable random
bits). The result is as follows.

\begin{fact} Every pure quantum formula computing a function $f$ with
  bounded error has length
\[\Omega\left(\sum_i D(f_i)/\log D(f_i)\right),\]
for the Ne\v ciporuk function $\sum_i D(f_i)$, see Fact 3.1 and
definition 3.4.
\end{fact}

Furthermore in \cite{RV01} it is shown that pure quantum formulae can
be simulated efficiently by deterministic circuits.

Now we know from \S 3.2 that the Boolean function $MP$ with
$O(n^2)$ inputs (the matrix product function) has fair probabilistic
formulae of linear  size $O(n^2)$, while the Ne\v ciporuk bound is cubic
(theorems 3.11 and 3.12). Thus we get the following.

\begin{corollary} There is a Boolean function $MP$ with $N$ inputs, which
  can be computed by fair Monte Carlo formulae of length $O(N)$,
  while every pure quantum formula with bounded error for $MP$ has size
$\Omega(N^{3/2}/\log N)$.
\end{corollary}

We conclude that pure quantum formulae are not a proper generalization
of classical formulae. A fair probabilistic formula can be simulated
efficiently by a generalized quantum formula on the other hand. We now
derive a lower bound method for generalized quantum formulae. First we
give again a lower bound in terms of one-way communication complexity,
then we show that the VC-Ne\v ciporuk bound is a lower bound, too.

This implies with Theorem 3.9 that the maximal difference between the
 sizes of deterministic formulae and generalized bounded error quantum
 formulae provable with the Ne\v ciporuk method is at most
 $O(\sqrt{n})$.

But first let us conclude the following corollary, which states that
fair probabilistic formulae reading their random bits only once are
sometimes inefficient.

\begin{corollary}
The (standard) Ne\v ciporuk function divided by $\log n$ is an
asymptotical lower bound for the size for fair probabilistic formulae
reading their random inputs only once.
\end{corollary}

\begin{proof} We have to show that pure quantum formulae can simulate
these special probabilistic formulae. For each random input we use two
qubits in the state $|00\rangle$. These are transformed into the state
$|\Phi^+\rangle$ by a Hadamard gate. One of the qubits is never used
again, then the other qubit has the density matrix of a random
bit. Then the probabilistic formula can be simulated. For the
simulation of gates unitary transformations on three qubits are
used. These get the usual inputs of the gate simulated plus one empty
qubit as input, which after the application of the gate carries the
output. These gates are easily constructed unitarily. According to
\cite{BBC95} each 3 qubits gate can be composed of  $O(1)$ unitary
gates on 2 qubits only.
\qquad\end{proof}

We will need the following observation \cite{AKN98}.

\begin{fact} If the density matrix of two qubits in a circuit (with
  nonentangled inputs) is not
  the tensor product of their density matrices, then there is a gate
  so that both qubits are reachable on a path from that gate.
\end{fact}

Since the above situation is impossible in a formula, the inputs to a
gate are never entangled.

The first lower bound is stated in terms of one-way communication
complexity. It is interesting that actually randomized complexity
suffices for a lower bound on quantum formulae.

\begin{theorem} Let $f$ be a Boolean function on $n$ inputs and
$y_1\ldots y_k$ a partition of the input variables in  $k$ blocks.
Player Bob knows the inputs in $y_i$ and player Alice knows all
other inputs. The randomized (private coin) one-way communication
complexity of $f$ (with bounded error) under this input partition
is called $R(f_i)$.

Every generalized
quantum formula for $f$ with bounded error has length
\[\Omega\left(\sum_i \frac{R (f_i)}{\log R(f_i)}\right).\]
\end{theorem}

\begin{proof} For a given partition of the input we show how a
generalized quantum formula $F$ can be simulated in the $k$
communication games, so that the randomized one-way communication
in game $i$ is bounded by a function of the number of leaves in a
subtree $F_i$ of $F$. $F_i$ contains exactly the variables
belonging to Bob as leaves and its root is the root of $F$.
Furthermore $F_i$ contains all gates on paths from these leaves to
the root. Note that the additional nonentangled mixed state which
the formula may access is given to Alice.

$F$ is a tree of fan-in 2 fan-out 1 superoperators (recall that
superoperators are not necessarily reversible). "Wires" between
the gates carry one qubit each. $F_i$ is a formula that Bob wants
to evaluate, the remaining parts of the formula $F$ belong to
Alice, and she can easily compute the density matrices for all
qubits on any wire in her part of the formula by a classical
computation, as well as the density matrices for the qubits
crossing to Bob's formula $F_i$. Note that none of the qubits on
wires crossing to $F_i$ is entangled with another, so the state of
these qubits is a probabilistic ensemble of pure nonentangled
states. Hence Alice may fix a pure nonentangled state from this
ensemble with a randomized choice.

In all communication games Bob evaluates the formula as far a
possible without the help of Alice. By an argument as in other
Ne\v ciporuk methods (e.g.~\cite{BS90,RV01} or the previous
sections) it is sufficient to send few bits from Alice to Bob to
evaluate a path with the following property: all gates on the path
have one input from Alice and one input from it predecessor,
except of the first gate, which has one input from Alice, and one
(already known) input from Bob. With standard arguments the number
of such paths is a lower bound on the number of leaves in the
subformula, see \S 3.1.

Hence we have to consider some path $g_1,\ldots, g_m$ in $F$,
where $g_1$ has one input or a gate from Alice as predecessor and
and input or gate from Bob as the other predecessor, and all gates
$g_i$ have the previous gate $g_{i-1}$ and an input or gate from
Alice's part of the formula as predecessors. The density matrix of
Bob's input to to $g_1$ is called $\rho$, and the density matrix
of the other $m$ inputs is called $\sigma$. The circuit computing
$\sigma$ works on different qubits than the circuit computing
$\rho$.

Thus the density matrix of all inputs to the path is
$\rho\otimes\sigma$, see Fact 6.8. The path maps
$\rho\otimes\sigma$ with a superoperator $T$ to a density matrix
$\mu$ on one qubit, altenatively we may view $\sigma$ as
determining a superoperator $T_\sigma$ on one qubit that has to be
applied to $\rho$. Now Alice can compute this superoperator by
herself, classically.

Bob knows $\rho$. Bob wants to know the state $T_\sigma \rho$.
Since this operator works on a single qubit only, it can be
described within a precision $1/poly(k)$ by a constant size matrix
containing numbers of size $O(\log k)$ for any integer $k$. Thus
Alice may communicate $T_\sigma$ to Bob within this precision
using $O(\log k)$ bits.

In this way Alice and Bob may evaluate the formula. and the error
of the formula is changed only by $size_i/poly(k)$ compared to the
error of the quantum formula, when $size_i$ denotes the number of
gates in $F_i$. Thus choosing $k=poly(size_i)$ the communication
is bounded $R (f_i)\le O(size_i\log size_i)$. This implies
$size_i\ge \Omega(R(f_i)/\log R(f_i))$. Summation over all $i$
yields the theorem. \qquad\end{proof}

The above construction loses a logarithmic factor, but in the
combinatorial bounds we actually apply, we can avoid this, by
using quantum communication and the programmable quantum gate from
Fact 4.20.

\begin{theorem}
The VC-Ne\v ciporuk function is an asymptotical lower bound for the length
of generalized quantum formulae with bounded error.

The Ne\v ciporuk function is an asymptotical lower bound for the length
of generalized quantum Las Vegas formulae.
\end{theorem}

\begin{proof} We proceed similar to the above construction, but
Alice and Bob use quantum computers. Instead of communicating a
superoperator in matrix form with some precision we use the
programmable quantum gate.

Alice and Bob cooperatively evaluate the formulae $F_i$ in a
communication game as before. As before, for certain paths Alice
wants to help Bob to apply a superoperator $T_\sigma$ on a state
$\rho$ of his. Using Kraus representations (Fact 4.7) we can
assume that this is a unitary operator on $O(1)$ qubits (one of
them $\rho$, the others blank) followed by throwing away all but
one of the qubits.

This time Alice sends to Bob the program corresponding to the
unitary operation in $T_\sigma$. Bob feeds this program into the
programmable quantum gate, which tries to apply the
transformation, and if this is successful the formula evaluation
can continue after discarding the unnecessary qubits. This happens
with probability $\Omega(1)$. If Alice could get some notification
from Bob saying whether the gate has operated successfully and if
not, what kind of error occurred, then Alice could send him
another program that both undoes the error and the previous
operator and then makes another attempt to compute the desired
operator.

Note that the error that resulted by an application of the
programmable quantum gate is determined by the classical
measurement outcome resulting in its application. Furthermore this
error can be described by a unitary transformation itself. If the
error function is $E$, the desired is unitary is $U$, and the
state it has to be applied to is $\rho$, then Bob now holds
$UE\rho E^\dagger U^\dagger$. Once Alice knows $E$ (which is
determined by Bob's measurement outcome), Alice can produce a
program for $U E^\dagger U^{\dagger}$. If Bob applies this
transformation successfully they are done, otherwise they can
iterate. Note that only an expected number of $O(1)$ such
iterations are necessary, and hence the expected quantum
communication in this process is $O(1)$, too.

So the expected communication can be reduced to $O(size_i)$. But
Alice needs some communication from Bob. Luckily this
communication does not reveal any information about Bob's input:
Bob's measurement outcomes are random numbers without correlation
with his input.

So we consider the nonstandard one-way communication model from
lemma 5.13, in which Bob may talk to Alice, but without revealing
any information about his input. Using this model in the
construction and letting Bob always ask explicitly for more
programs reduces the communication in game $i$ to $O(size_i)$ in
the expected sense.

With lemma 5.13 we get the lower bounds for bounded error and Las
Vegas communication.
\qquad\end{proof}

Now we can give a lower bound for $ISA$ showing that even generalized
quantum formulae compute the function not significantly more efficient
than deterministic formulae.

\begin{corollary}
Every generalized quantum formula, which computes $ISA$ with bounded
error has length $\Omega(n^2/\log n)$.
\end{corollary}

Considering the matrix multiplication function $MP$ we get the following.

\begin{corollary} There is a function, which can be computed by a
  generalized quantum formula with bounded error as well as by a fair
  probabilistic formula with bounded error, with size $O(N)$. Every
  generalized quantum Las Vegas formula needs size
$\Omega(N^{3/2})$ for this task. Hence there is a size gap of
  $\Omega(N^{1/2})$ between
Las Vegas formula length and the length of bounded error formulae.
\end{corollary}

Since the VC-Ne\v ciporuk function is a lower bound for
generalized quantum formulae, Theorem 3.9 implies that the maximal
size gap between deterministic formulae and generalized quantum
formulae with bounded error provable by the (standard) Ne\v
ciporuk method is $O(\sqrt{n})$ for input length $n$. Such a gap
actually already lies between generalized quantum Las Vegas
formulae and fair probabilistic formulae with bounded error.

\section{Conclusions}

In this paper we have derived lower bounds for the sizes of
probabilistic, nondeterministic, and quantum formulae. These lower
bounds follow the general approach of reinterpreting the Ne\v
ciporuk bound in terms of one-way communication complexity. This
is nontrivial in the case of quantum formulae, where we had use a
programmable quantum gate. Nevertheless we have obtained the same
combinatorial lower bound for quantum and probabilistic formulae
based on the VC-dimension.

Using the lower bound methods we have derived a general $\sqrt{n}$
gap between bounded error and Las Vegas formula size. Another result
is a threshold phenomenon for the amount of nondeterminism needed to
compute a function, which gives a near -quadratic size gap for a
polynomial threshold on the number of nondeterministic bits.

To derive our results we needed lower bounds for one-way
communication complexity. While these were available in the case
of probabilistic one-way communication complexity, we had to
develop these lower bounds in the quantum and nondeterministic
case. These results give gaps between 2-round and one-way
communication complexity in these models. Those gaps have been
generalized to round hierarchies for larger number of rounds in
\cite{Kl98} and \cite{KNTZ01} for the nondeterministic resp.~the
quantum case. Furthermore we have shown that quantum Las Vegas
one-way protocols for total functions are not much more efficient
than deterministic one-way protocols. The lower bounds for quantum
one-way communication complexity are also useful to give lower
bounds for quantum automata, and for establishing that only
bounded error quantum finite automata can be exponentially smaller
than deterministic finite automata \cite{Kl00}. A generalization
of the VC-dimension bound on quantum one-way communication
complexity is given in \cite{Kl00b}.

We single out the following open problems:
\begin{remunerate}
\item Give a better separation between deterministic and
probabilistic/quantum formula size (see \cite{Kl97} for a
candidate function).
\item Separate the size complexities of
generalized quantum and probabilistic formulae for some function.
\item Investigate the power of quantum formulae that can access an entangled
state as an additional input, thus introducing entanglement into the model.
\item Separate quantum and probabilistic one-way communication
  complexity for some total function or show that both are related.
\item Prove super-quadratic lower bounds for formulae over the basis
  of all two-ary Boolean functions.
\end{remunerate}

\section*{Acknowledgments}

The author wishes to thank Gregor Gramlich for a talk on the Ne\v
ciporuk method, which inspired this research, and Georg Schnitger
for stimulating discussions.

\end{document}